\documentclass[11pt]{article}

\usepackage{packages}
\usepackage{statistics-macros}
\usepackage{formatting}
\usepackage{booktabs}
\usepackage{microtype}
\usepackage{mathtools}
\usepackage{array}
\usepackage{multirow}
\usepackage{placeins}
\usepackage[capitalize,noabbrev]{cleveref}

\crefname{assumption}{Assumption}{Assumptions}
\Crefname{assumption}{Assumption}{Assumptions}
\crefname{sensitivityrestriction}{Restriction}{Restrictions}
\Crefname{sensitivityrestriction}{Restriction}{Restrictions}
\crefname{observation}{Observation}{Observations}
\Crefname{observation}{Observation}{Observations}

\makeatletter
\renewcommand*{\theHALG@line}{\thealgorithm.\arabic{ALG@line}}
\makeatother

\numberwithin{equation}{section}

\graphicspath{{figures/}}
\hypersetup{
  pdftitle={A Sensitivity Framework for Assessing External Validity},
  pdfauthor={Ariel Boyarsky, Naoki Egami, and Hongseok Namkoong},
  pdfborder={0 0 0}
}

\makeatletter
\@ifundefined{remark}{\newtheorem{remark}{Remark}}{}
\makeatother

\newcommand{\tP}{Q_T}                    %

\newcommand{\ct}{c_T}                    %
\newcommand{\muC}{m_C}                   %
\newcommand{\Ctil}{\widetilde C}         %
\newcommand{\propen}[1]{e_{#1}}          %
\newcommand{\pseudo}[1]{U_{#1}}          %
\newcommand{\zetaratio}{\zeta}           %

\newcommand{\CovC}{\Sigma_C}             %
\newcommand{\dmom}[1]{d_{#1}}            %
\newcommand{\bT}{b_T}                    %
\newcommand{\ellvec}{\ell_T}             %

\newcommand{\CovR}[1]{\E_P[\cov(C,R_{#1}\mid X)]}
\newcommand{\CovRdiff}{\E_P[\cov(C,R_1-R_0\mid X)]}
\newcommand{\whatCovC}{\what\E[\what\cov(C\mid X)]}
\newcommand{\whatCovR}[1]{\what\E[\what\cov(C,R_{#1}\mid X)]}

\newcommand{\Vspan}{\mathcal V_C}
\newcommand{\Bspan}{B_C}
\newcommand{\Om}{\Omega}                 %
\newcommand{\xired}[1]{\varsigma_{#1}}         %
\newcommand{\dxi}[1]{\dot\varsigma_{#1}}       %
\newcommand{\gred}{\gamma}               %
\newcommand{\Hred}{H}                    %
\newcommand{\ered}{e}                    %
\newcommand{\redomain}{\mathcal U}       %

\newcommand{\restrset}{\mathcal R}
\newcommand{\feasset}{\mathcal F}
\newcommand{\branchval}{u}               %
\newcommand{\vminus}{v_{-}}
\newcommand{\vplus}{v_{+}}
\newcommand{\lamred}{\bar\lambda}        %
\newcommand{\nuineq}{\nu}                %
\newcommand{\kkt}{\xi}                   %
\newcommand{\kappaOm}{\kappa_\Omega}          %

\DeclareMathOperator{\rankop}{rank}
\DeclareMathOperator{\rangeop}{range}
\DeclareMathOperator{\kernelop}{ker}

\DeclareMathOperator{\affop}{aff}
\DeclareMathOperator{\convop}{conv}
\newcommand{\range}{\rangeop}
\newcommand{\kernel}{\kernelop}
\newcommand{\pr}{\mathbb{P}}
\newcommand{\one}{\mathbbm{1}}

\newcommand{\Pnk}[1]{\mathbb P_{n,#1}}

\newcommand{\Eone}{\textsc{pd}}
\newcommand{\Etwo}{\textsc{os}}
\newcommand{\vhatE}[2]{\what v^{#1}_{#2}}     %
\newcommand{\gplugk}[1]{\what\gred^{\,\mathrm{pl},(#1)}}

\makeatletter
\@ifundefined{proofof}{%
  }{}
\makeatother

\begin{document}

\begin{center}
  {\LARGE A Sensitivity Framework for Assessing External Validity}\\[0.2cm]
  {\large Ariel Boyarsky$^{*}$ \qquad Naoki Egami$^{\dagger}$ \qquad Hongseok Namkoong$^{*}$}\\[0.2cm]
  {$^{*}$Decision, Risk, and Operations Division, Columbia Business School\\
   $^{\dagger}$Department of Political Science, Statistics and Data Science Center,\\ Massachusetts Institute of Technology}\\[0.2cm]
  \texttt{a.boyarsky@columbia.edu, egami@mit.edu, namkoong@gsb.columbia.edu}
\end{center}

\begin{abstract}
Assessing the external validity of experimental results is a pervasive
challenge in causal inference as internally valid findings need not
generalize beyond sites in which the experiments were conducted. Since the number of sites is often smaller than the number of site-level variables that moderate causal effects, we develop a partial identification framework that lets practitioners assess external
validity under a wide range of assumptions that allow extrapolation across site-level variables. Our optimization-driven approach can accommodate any
affine family of parameter restrictions, such as interaction terms being dominated by direct effects (the ``effect hierarchy principle''). Since standard estimating-equation arguments do not apply in our setting---underspecified parameter directions responsible for partial identification make the covariance matrix singular---we develop a novel debiasing approach for estimands given by constrained optimization problems, giving rise to semiparametrically efficient estimators for our flexible partial identification bounds.  We illustrate our approach on multi-site microcredit experiments harmonized by \citet{meager2019understanding}, whose lending environments differ widely in
interest rates, loan sizes, and promotion. Holding out one country at a
time, we construct external-validity intervals for the held-out country from
the remaining source sites and show how our framework connects different
extrapolation assumptions to concrete partial identification bounds.

\end{abstract}

\section{Introduction}
\label{sec:introduction}

Randomized experiments are conducted in particular places and times, under
particular implementations. The settings in which their evidence is used
rarely match the settings in which it was produced: institutions, program
terms, and local economic conditions all change when an intervention moves
\citep{bracht1968external,deaton2018understanding}. When treatment effects
vary with these site-level variables, internally valid findings lose external
validity outside their source sites~\citep{allcott2015site,vivalt2020generalize,meager2019understanding}.  A
mature set of methods addresses compositional differences between
sites, i.e., differences in characteristics of individuals in each (X-validity
in~\citet{egami2023elements}'s taxonomy), by holding the conditional treatment
effect fixed and adjusting for population compositions
\citep{stuart2011propensity,tipton2013improving,dahabreh2019generalizing,degtiar2023review}. In
this paper, our focus is on differences in causal effects inherent to
site-level variables (C-validity in~\citet{egami2023elements}'s taxonomy). 

Consider a policymaker using microcredit experiments in six countries to
predict the effect on household revenues in Bosnia and Herzegovina, our main
application in \cref{sec:case-study}. Interest rates, loan sizes, and promotion
differ across countries in the data of \citet{meager2019understanding}.
To what extent can the source trials predict the effect in Bosnia when its
lending environment differs from those in which the trials were conducted?

We propose a sensitivity framework in which site-level variables directly affect causal effects beyond individual-level variables. The fundamental
problem is that there are often fewer source sites than site-level variables to estimate and thus the effects of site-level
variables are only partially identified.\footnote{This limitation is separate from overlap in individual
  covariates. Even with the same population composition as the source sites,
  a target effect can depend on the effects of site-level variables that the source data do
  not identify.} In the example
above, there are only six sites but scholars have pointed out more
than six site-level variables that can moderate causal effects across
sites. In such cases, even if the average treatment effect
in each source site were known exactly, the average treatment effect
in a new site is often not point-identified. 

To address this fundamental identification problem, we consider a
partially linear model of the potential outcomes, which we formally
introduce in the subsequent sections. For $a\in\{0,1\}$, we have 
\begin{equation}
  \E_P[Y(a) \mid X, C] = g_a(X) + C^\top\beta_a,
\end{equation}
where $Y(a)$ is the potential outcome, $X$ are the unit-level
variables, $C$ are the site-level variables, and $\E_P$ denotes the
expectation over the source population distribution. $g_a$ is an unrestricted function that captures complex unit-level
heterogeneity, and $\beta_a\in\R^p$ is a fixed vector of coefficients
for site-level variables. Note that
unit-level variables $X$, such as age and gender, can vary within
each site, but site-level variables $C$, such as local economic and
political conditions, can differ only across sites.  Given this model, we can express the average
treatment effect in a new target site as $\psi_T = \E_{X\sim Q_T}[g_1(X)-g_0(X)] + c_T^\top(\beta_1-\beta_0)$
where $Q_T$ denotes the target population distribution of $X$
and $c_T$ is a vector of site-level variables in the
particular target site.  

Even though this partially linear model is simple, when the number of
sites is smaller than the number of site-level variables $C$, as is
often the case, the coefficients $\beta_a$ are typically not identified, which often
implies that the average treatment effect in the target site $\psi_T$ is also not identified. For
partial identification, we retain all coefficient vectors consistent with the source evidence and
impose explicit restrictions on their magnitudes, signs, or the size of
higher-order terms relative to direct effects of site-level variables (\cref{sec:small-sites}).
Minimizing and maximizing the target effect over these vectors gives
partial identification bounds. Varying the restrictions shows how
different assumptions affect partial identification bounds. Wide partial identification bounds indicate that substantial
uncertainty remains even under the stated restrictions; an unbounded endpoint
means that some extrapolations remain unrestricted. An empty feasible set
means that the restrictions contradict the source evidence. 

The unit-lvel heterogeneity component $g_a(X)$ introduces a further challenge because differences
in population composition can confound comparisons across sites. We
allow outcomes to depend arbitrarily on unit-level covariates $X$, provided
this dependence remains unchanged across sites within each treatment arm,
and adjust for the target distribution of $X$. Partialling out $X$ using
the approach of \citet{robinson1988root} reduces the source model to a
linear system for each arm (\cref{sec:framework}). The resulting matrix is
singular when the source sites do not identify all the effects of
site-level variables $\beta_a$. Errors from regressing site-level variables $C$ on unit-level covariates $X$
can nevertheless make the estimated matrix full rank and collapse the
feasible set to a point. We develop an estimator that preserves the unidentified coefficient combinations, so regression error
cannot create spurious point identification (\cref{sec:estimation}).

We develop a debiasing procedure for inference on these partial identification bounds. The
theoretical challenge is that each partial identification bound solves a constrained optimization
problem whose objective and equality constraints must both be estimated.
Estimation error changes the feasible set itself, and because the constraint
matrix is singular, arbitrarily small perturbations can make the equalities
infeasible. The usual smoothness argument behind debiasing therefore does
not apply directly. We overcome this obstacle by proving a first-order
expansion for the optimized partial identification bound under our estimator
(\cref{sec:asymptotics}). This result establishes that correcting bias in the
objective and constraints before optimization removes the first-order effect
of nuisance estimation error on the partial identification bounds. Under our regularity conditions,
we obtain valid confidence intervals while allowing flexible
estimation of unit-level heterogeneity.

We use the microcredit experiments to examine how modeling choices affect
conclusions about a new lending environment
(\cref{sec:case-study}). Comparing alternative restrictions shows which
assumptions are needed to determine the sign or magnitude of the target
effect. The same restriction can substantially narrow the partial identification bounds for one
target while leaving another largely unresolved, depending on the contexts
represented in the source experiments. Varying the target's lending terms
and removing source sites then distinguishes sensitivity to the target
setting from dependence on particular source experiments.

\paragraph{Related work.}
Methods for generalizing experiments identify target effects by adjusting for
population composition under a stable conditional treatment response and
covariate overlap \citep{dahabreh2019generalizing,
  degtiar2023review, egami2021covariate, jin2025beyond}, including
when several trials supply the source evidence \citep{dahabreh2023multiple}.
Our target covariate adjustment uses the same logic, but we also allow
site-level variables to change the response. Such heterogeneity is documented in field
experiments \citep{allcott2015site,meager2019understanding} and corresponds to
contextual validity in \citet{egami2023elements}'s taxonomy. We study what
source experiments imply when they do not identify all site-level effects
needed to predict the target effect. Using these partial identification bounds to choose a treatment
requires a further decision criterion, as in work on treatment choice under
partial identification and limited external validity
\citep{stoye2012minimax,montiel2026decision}. Our analysis characterizes the
range of target effects under stated restrictions; a policy decision also
requires a welfare criterion and a stance toward ambiguity.

Sensitivity analyses for external validity relax transportability assumptions by allowing omitted
effect modifiers \citep{nguyen2017unobserved, huang2024sensitivity} or explicit
deviations from exchangeability across study participation
\citep{dahabreh2023bias}. Other approaches assess external validity by imposing
covariate-balancing constraints under unmeasured effect modification
\citep{nie2021covariate}, restricting differences between source and target
outcome distributions \citep{asiaee2026sharp}, allowing for the
violation of the overlap assumption \citep{huang2025overlap}, or evaluating worst-case
treatment effects over subpopulations \citep{jeong2020assessing}.
We vary coefficients governing how site context changes
treated and control outcomes, while allowing heterogeneous
individual-level treatment effects that
remain unchanged across sites. Following the partial-identification approach
of \citet{Manski2003}, we retain every coefficient vector consistent with the
source evidence and the stated restrictions. Multiple experiments constrain
these coefficients, and restrictions on their signs and magnitudes limit the
remaining target extrapolations. Our sensitivity analysis describes the
underidentification that remains under the additive response model; it does
not assess violations of that model.

\section{Extrapolation with a small number of sites}
\label{sec:small-sites}

We begin with a warm-up in which there is no unit-level covariate $X$ and
the source-site mean outcomes are known. This separates the modeling and
identification questions from sampling uncertainty. Our framework then
reduces to a regression across sites, with potentially more coefficients than
the source experiments can identify. We first formulate the target partial identification bounds,
then describe how the analyst's choice of site-level variables and parameter
restrictions determines what can be learned. A two-site example illustrates
how these choices affect the partial identification bounds. We introduce the unit-level variable
$X$ in \cref{sec:framework} and address estimation from finite samples in
\cref{sec:estimation}.

\subsection{A partial-identification framework}

In the microcredit example, we observe experiments in a small number of
countries whose lending environments differ in interest rates, loan sizes,
and promotion. Let $c_s\in\R^p$ describe the site-level variables of source site
$s=1,\ldots,S$, and let $\ct\in\R^p$ describe the target site. The
source site-level vector $C$ takes one of the values $c_1,\ldots,c_S$. We choose
this site-level basis before the analysis; it may include raw characteristics,
their squares, and their interactions. Without a unit-level covariate $X$,
randomization and consistency identify each source potential-outcome mean
$m_{a,s}=\E[Y(a)\mid C=c_s]$, where $Y(a)$ is the outcome under treatment
arm $a\in\{0,1\}$.

We assume that site-level variables enter each potential-outcome mean linearly,
\begin{equation}
  m_{a,s}=\alpha_a+c_s^\top\beta_a,
  \qquad a\in\{0,1\},\quad s=1,\ldots,S,
  \label{eq:no-x-response-model}
\end{equation}
and that the same intercepts and coefficients govern the target means.
Subtracting the mean at source site 1 removes the intercept and gives
\begin{equation}
  (c_s-c_1)^\top\beta_a=m_{a,s}-m_{a,1},
  \qquad s=2,\ldots,S,\quad a\in\{0,1\}.
  \label{eq:no-x-source-equations}
\end{equation}
These are $S-1$ linear equations for each vector of $p$ coefficients. If
$p>S-1$, some coefficient combinations remain unidentified even when every
source mean is known. In the microcredit analysis for Bosnia, six source
sites supply at most five independent comparisons for nine site-level terms.
At least four linear combinations of each site-level vector are therefore
undetermined. Additional observations within each source site improve the precision of each source
mean, but cannot supply another comparison across site-level variables. 

Under \eqref{eq:no-x-response-model}, the target average treatment effect is
\begin{equation}
  \psi_T =m_{1,1}-m_{0,1}+(\ct-c_1)^\top(\beta_1-\beta_0).
  \label{eq:no-x-target-effect}
\end{equation}
The source sites identify this effect if $\ct-c_1$ is a linear combination
of the site-level differences $\{c_s-c_1:s=2,\ldots,S\}$. Otherwise,
coefficients that fit the same source means can imply different target
effects. We limit these extrapolations by imposing substantive restrictions
on the coefficients.

Let $\mathcal R$ contain the coefficient vectors
$\beta=(\beta_0^\top,\beta_1^\top)^\top$ allowed by the analyst's
restrictions. The vectors consistent with both the restrictions and
the source means form the feasible set
\begin{equation}
  \mathcal F^{(0)}
  =\left\{\beta\in\mathcal R:
    (c_s-c_1)^\top\beta_a=m_{a,s}-m_{a,1},\;
    s=2,\ldots,S,\; a\in\{0,1\}\right\}.
  \label{eq:no-x-feasible-set}
\end{equation}
The extreme target effects over this set are the
\emph{partial identification bounds},
\begin{equation}
\begin{split}
  v_-^{(0)}&=m_{1,1}-m_{0,1}
    +\inf_{\beta\in\mathcal F^{(0)}}
      (\ct-c_1)^\top(\beta_1-\beta_0),\\
  v_+^{(0)}&=m_{1,1}-m_{0,1}
    +\sup_{\beta\in\mathcal F^{(0)}}
      (\ct-c_1)^\top(\beta_1-\beta_0).
\end{split}
\label{eq:no-x-endpoint-programs}
\end{equation}
The optimizing coefficients give the extreme target effects allowed by the
data and restrictions. Relaxing a restriction can only widen the partial identification bounds.
Tightening it narrows the range of possible effects, but may eventually
exclude every coefficient vector consistent with the source evidence.

\subsection{Choosing parameter restrictions}
\label{subsec:examples}

The site-level basis determines which features of the site environment can
change outcomes. Parameter restrictions determine how much those features
can matter. Their interpretation depends on the scale of the
site-level variables, which we fix before choosing the restrictions.

\paragraph{Example 1: Magnitude caps.}
A magnitude cap imposes $|\beta_{a,j}|\leq M_j$ for chosen constants $M_j$.
In the microcredit application, we scale each site-level variable by its standard
deviation across source sites. The cap then limits the revenue change
attributed to a one-standard-deviation change in that term, holding the others
fixed. We set the cap to one standard deviation of the outcome and can vary
it to examine how the target partial identification bounds depend on the allowed effect sizes.

\paragraph{Example 2: Effect hierarchy principle.}
The effect hierarchy principle from the design of experiments, also called
effect ordering or hierarchical ordering, states that lower-order terms are
more likely to matter than higher-order terms
\citep{hamada1992analysis,montgomery2004heredity}. It is widely used in
statistics, for example in hierarchical interaction models
\citep{bien2013lasso,lim2015learning}. We use a stronger
restriction that compares the magnitude of every higher-order coefficient
with every direct coefficient. Write $C=(C_D^\top,C_H^\top)^\top$ for the
$p_D$ direct terms and $p_H$ higher-order terms, and partition each $\beta_a$
accordingly. For a sensitivity level $\kappa\geq0$,
\begin{equation}
  |\beta_{a,H,j}|\leq \kappa |\beta_{a,D,k}|
  \quad\text{for }a\in\{0,1\},\ j=1,\ldots,p_H,\ k=1,\ldots,p_D.
  \label{eq:effect-ordering}
\end{equation}

At $\kappa=0$, all higher-order coefficients are zero. At $\kappa=1$, each
is no larger in magnitude than any direct coefficient in the same treatment
arm. Larger values relax the comparison. A direct coefficient near zero
therefore forces every higher-order coefficient toward zero, so this can be
a strong restriction. Varying $\kappa$ shows how much the target conclusion
depends on it.

\paragraph{Example 3: Aggregate budgets.}
An aggregate budget instead limits the sum of higher-order coefficient
magnitudes relative to the sum for direct coefficients.
\begin{equation}
  \sum_{j=1}^{p_H}|\beta_{a,H,j}|
  \;\leq\;\kappa\sum_{k=1}^{p_D}|\beta_{a,D,k}|,
  \qquad a\in\{0,1\}.
  \label{eq:l1-budget}
\end{equation}
This permits one large higher-order coefficient when the others are small,
whereas \eqref{eq:effect-ordering} limits each coefficient separately.

\paragraph{Example 4: Sign and shape restrictions.}
Subject-matter knowledge can also restrict the sign of particular
coefficients. For example, in the microcredit application, a nonpositive direct interest-rate
coefficient in the treatment effect gives
\begin{equation}
  \beta_{1,\mathrm{APR}}-\beta_{0,\mathrm{APR}}\leq0.
  \label{eq:apr-sign}
\end{equation}
An analyst could also require  effects of site-level variables to increase or decrease
across ordered categories, such as discretized interest rates.

We recommend adding restrictions one at a time and reporting how each changes
the target partial identification bounds. Varying the sensitivity parameters within each family
shows how the conclusion depends on the strength of the restriction. We
follow this protocol in \cref{sec:case-study}.

\subsection{Numerical example}
\label{subsec:numerical-example}

\begin{figure}[t]
  \centering
  \begin{tikzpicture}[scale=2.1, >=stealth, every node/.style={font=\small}]
  \fill[gray!12] (-1,-1) rectangle (1,1);
  \draw[gray!70] (-1,-1) rectangle (1,1);
  \node[gray!50!black] at (-0.58,-0.84) {$|\theta_j|\le M$};
  \draw[->] (-1.6,0) -- (2.15,0) node[below] {$\theta_1$};
  \draw[->] (0,-1.55) -- (0,1.95) node[left] {$\theta_2$};
  \draw (-1,-0.03) -- (-1,0.03);
  \node[below] at (-1,-0.05) {$-M$};
  \draw (1.2,-0.03) -- (1.2,0.03);
  \node[below right] at (1.17,-0.03) {$\rho$};
  \draw[densely dotted, green!45!black] (-1.35,1.65) -- (1.55,-1.25);
  \node[green!45!black, rotate=-45, anchor=south] at (-0.98,1.26)
    {\scriptsize level set of the $(2,2)$ objective};
  \draw[blue!70!black, thick] (-0.35,1.55) -- (1.95,-0.75);
  \node[blue!70!black, rotate=-45, anchor=south] at (1.75,-0.57)
    {\scriptsize $\theta_1+\theta_2=\rho$};
  \draw[<->, thick] (-0.52,1.92) -- (0.08,1.32);
  \node[anchor=west] at (0.12,1.66) {\scriptsize unidentified direction $(1,-1)$};
  \draw[orange!90!black, line width=1.8pt] (0.2,1) -- (1,0.2);
  \draw[orange!80!black, ->, thin] (1.32,0.92) -- (0.72,0.52);
  \node[orange!80!black, anchor=south] at (1.32,0.92) {\scriptsize feasible set};
  \draw[dashed, gray] (0.2,1) -- (0.2,0);
  \draw[dashed, gray] (1,0.2) -- (1,0);
  \draw[line width=1.6pt] (0.2,0) -- (1,0);
  \node[below] at (0.2,-0.05) {\scriptsize $\rho{-}M$};
  \node[below] at (1,-0.05) {\scriptsize $M$};
  \node[anchor=north] at (0.6,-0.3) {\scriptsize interval for $\tau(1,0)-\tau_0$};
\end{tikzpicture}
  \caption{Partial identification with two source sites and no unit-level
    covariate $X$. The blue line contains the coefficients consistent with
    the source effects. The gray box imposes the magnitude cap, leaving the
    orange feasible segment. Its projection onto $\theta_1$ gives the partial identification bounds
    for the target at $(1,0)$ after adding $\tau_0$. Here $\rho=1.2$ and $M=1$.}
  \label{fig:two-sites}
\end{figure}
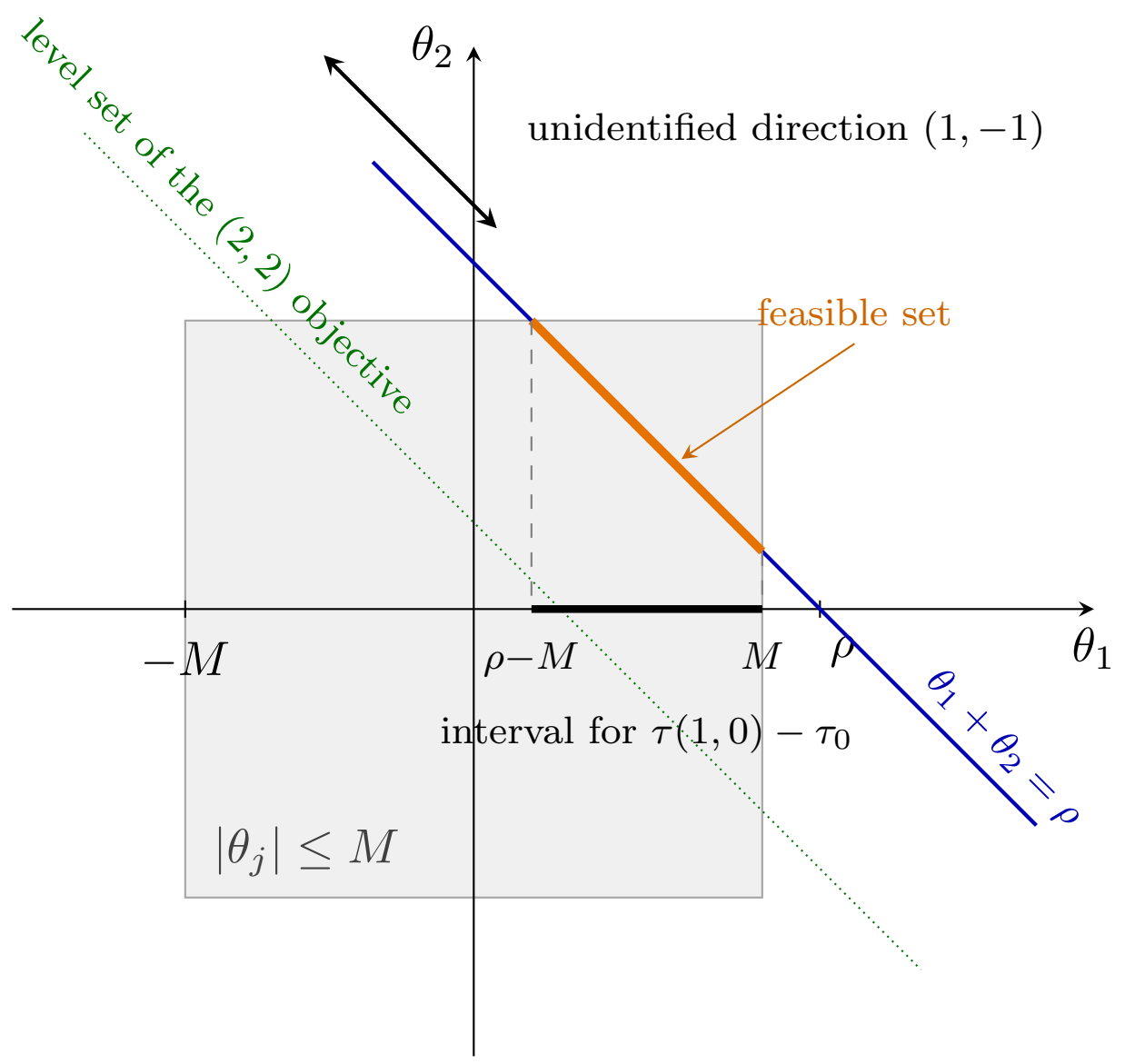
\FloatBarrier

We give a minimal example to provide mathematical intuition for how
the source site-level variables, target site-level variables, and parameter restrictions together determine the
partial identification bounds.  Consider the treatment-effect surface
\begin{equation}
  \tau(c)=\tau_0+c_1\theta_1+c_2\theta_2,
  \qquad \theta_j=\beta_{1,j}-\beta_{0,j},\quad j\in\{1,2\},
  \label{eq:two-site-effect-surface}
\end{equation}
where $j$ indexes the two site-level variables, and $c_1$ and $c_2$ are the coordinates of the site-level vector in this
example. Source sites $(0,0)$ and $(1,1)$ identify $\tau_0$ and the sum
$\rho=\theta_1+\theta_2$. Every coefficient pair on the blue line in \cref{fig:two-sites}
therefore fits both source effects. At target site $(2,2)$, the effect is $\tau_0+2\rho$ and is point
identified. At target site $(1,0)$, the effect is $\tau_0+\theta_1$, which varies along the blue line. The same source
evidence can therefore identify the effect at one target and leave it undetermined at another.

Impose a magnitude cap $|\theta_j|\leq M$ directly on the treatment-effect
coefficients. This leaves the orange segment and restricts the effect at
$(1,0)$ by
\begin{equation}
  \left[\tau_0+\max\{-M,\rho-M\},\;
        \tau_0+\min\{M,\rho+M\}\right],
  \qquad |\rho|\leq 2M.
  \label{eq:two-site-bounds}
\end{equation}
For $\rho=1.2$ and $M=1$, as shown in \cref{fig:two-sites}, the partial identification bounds are
$[\tau_0+0.2,\,\tau_0+1]$. The interval tightens as $M$ falls. If
$|\rho|>2M$, no coefficient pair satisfies both the cap and the source evidence.

\section{A partial-identification framework with unit-level covariates}
\label{sec:framework}

We now introduce unit-level covariates $X$. The model in
\cref{sec:small-sites} allowed site-level variables to explain differences in source
means. Here we separate site-level effects from differences in population
composition by allowing an unrestricted response to $X$ in each treatment
arm. Partialling out these responses gives a linear system analogous to
\eqref{eq:no-x-source-equations}. We characterize which target effects this
system identifies and combine it with the parameter restrictions from
\cref{subsec:examples}.

We denote our observations $O = (Y, A, X, C)$, where $A\in\{0,1\}$ is treatment,
$X$ contains unit-level covariates, and $C\in\R^p$ is the site-level vector shared by every observation in the same site. We choose the basis of
$C$ before the analysis; it may contain raw site characteristics and their
transformations. The source distribution $P$ pools the $S$ sites, and within
each site, we assume treatment is randomized with known probability
$e_a(x,c)=\pr_P(A=a\mid X=x,C=c)$, bounded away from zero.  The target is
described by a site-level profile $\ct\in\R^p$ and a distribution $X \sim Q_T$ of
unit-level covariates, from which a sample is available. This sample can come
from a census, survey, or even a prior over the unit-level covariates. Treatment and outcome data come only from the source (observed)
sites.\footnote{The pair $(Q_T,\ct)$ can be read as a conditional of one
  extended population: adjoin a target site to the $S$ source sites and
  define the covariate law at that site to be $Q_T$ with target
  site-level variable $\ct$,
  leaving the source conditionals $P(X\in\cdot\mid C=c_s)$ unchanged. Every
  target-subscripted object is then a conditional of this single law at the
  target site. We keep the explicit pair for two reasons. First, the
  conditioning must be on a site label rather than on the site-level value:
  $P(\,\cdot\mid C=\ct)$ is undefined when $\ct$ lies outside the source
  support, and when $\ct$ coincides with an observed profile it describes
  units already in the source sites, whose effect is point identified
  \eqref{eq:id-target-affine-span}, not a new location. Second, the two
  samples are drawn separately with fixed sizes, and the two-sample notation
  tracks this in the asymptotic theory.}

Randomization lets us express potential-outcome means as regressions of an
observable inverse-probability pseudo-outcome. Define
\begin{equation}
  R_a=\frac{\one\{A=a\}Y}{e_a(X,C)}, \qquad a\in\{0,1\}.
  \label{eq:pseudo-outcome}
\end{equation}
Because treatment is randomized given $(X,C)$, the pseudo-outcome satisfies
$\E_P[R_a\mid X,C]=\E_P[Y(a)\mid X,C]$. This standard transformation does not
resolve cross-site extrapolation since $\E_P[Y(a)\mid X=x,C=c]$ may be
unidentified for certain values of $c$.

\subsection{A structured response model}
\label{subsec:model}

Without a cross-site restriction, the conditional mean at a target site
outside the source support can vary arbitrarily without changing the source
distribution. We therefore impose an additive structure across sites.
Site-level variables enter each potential-outcome surface through linear fixed
effects, while heterogeneity in unit-level covariates remains unrestricted.
\begin{assumption}[A Partially Linear Potential Outcome Model with Site-level Effects]
  \label{ass:model}
  \label{eq:app-response-model}
For $a\in\{0,1\}$,
\begin{equation}
  \E_P[Y(a)\mid X,C]=g_a(X)+C^\top\beta_a,
  \label{eq:response-model}
\end{equation}
where $g_a$ is unrestricted and $\beta_a\in\R^p$ is a fixed vector of
site-level fixed effects. Treatment is randomized within each source
site, consistency holds, and the same functions and coefficients govern
the target potential-outcome means at $(Q_T,\ct)$. We assume $\tP\ll P_X$
and write $w=d\tP/dP_X$. All second moments below are finite.
\end{assumption}
\noindent With no $X$, each $g_a$ is a constant $\alpha_a$, recovering
\eqref{eq:no-x-response-model}.

In the microcredit application, $g_a(X)$ allows revenues under arm $a$ to vary
arbitrarily with household age, education, business ownership, and the other
unit-level covariates. The site-level variables $C$ allow APR, loan size, and promotion
to shift each arm's average outcome by a common amount across
households. The model excludes
interactions between unit-level covariates and site context. For example, it
cannot represent an APR effect that differs with prior business ownership, and
adding terms to the basis of $C$ does not relax this exclusion.  Our
sensitivity analysis quantifies the underidentification that remains under
\cref{ass:model}, and cannot assess violations of the additive response
structure.

Averaging the response model over the target population gives the target
average treatment effect,
\begin{equation}
  \psi_T =\E_{X\sim Q_T}[g_1(X)-g_0(X)]+\ct^\top(\beta_1-\beta_0).
  \label{eq:target-objective}
\end{equation}
To express this effect in terms of observable source regressions, take
conditional expectations given $X$ in
$\E_P[R_a\mid X,C]=g_a(X)+C^\top\beta_a$. This gives
$g_a(x)=\E_P[R_a\mid X=x]-\E_P[C\mid X=x]^\top\beta_a$ and therefore
\begin{equation}
  \psi_T=b_T+\ellvec^\top(\beta_1-\beta_0),
  \qquad
  \begin{aligned}
    b_T&=\E_{X\sim Q_T}\bigl[\E_P[R_1-R_0\mid X]\bigr],\\
    \ellvec&=\ct-\E_{X\sim Q_T}\bigl[\E_P[C\mid X]\bigr].
  \end{aligned}
  \label{eq:target-coefficients}
\end{equation}
\label{eq:app-target-coefficients}
\label{eq:app-target-effect}
Equation \eqref{eq:target-coefficients} has the same form as
\eqref{eq:no-x-target-effect}, with the baseline and site-level difference now
adjusted for population composition. The term $b_T$ averages the source
regression under the target covariate law and is identified under overlap
in $X$. The site-level difference $\ellvec$ measures how the target
site differs from the source site predicted for target-like units. It
multiplies the fixed-effect differences $\beta_1-\beta_0$, the only unknown quantity in
\eqref{eq:target-coefficients}. Whether $\psi_T$ is point identified therefore
depends on whether the source sites identify the direction of $\ellvec$.

\subsection{What the source sites identify}
\label{subsec:identification}

Under \cref{ass:model}, $\beta_a$ is the coefficient on $C$ in a partially
linear regression problem. The observable regression
\begin{equation}
  \E_P[R_a\mid X,C]=g_a(X)+C^\top\beta_a
  \label{eq:robinson-model}
\end{equation}
has $R_a$ as its outcome, $C$ as the low-dimensional regressor of interest,
and an unrestricted function $g_a$ of the unit-level covariates. Its
population minimizers solve
\begin{equation}
  \min_{f,\,\beta}\
  \E_P\bigl[\{R_a-f(X)-C^\top\beta\}^2\bigr].
\end{equation}
\Citet{robinson1988root} profiles out $f$ by residualizing both $R_a$ and $C$
on $X$. The normal equations are
\begin{equation}
  \CovC\,\beta_a
  =\CovR{a},
  \qquad a\in\{0,1\},
  \label{eq:profiled-moments}
\end{equation}
where $\CovC=\E_P[\cov(C\mid X)]$ measures the variation of site-level
variables remaining after adjustment for $X$. Since $C$ takes only $S$ values, its residuals lie
in the span of the $S-1$ centered site profiles, so
$\rankop(\CovC)\leq S-1$. Sampling more units cannot identify the
missing directions in site-level variables.

The coefficient vectors consistent with the source conditional means are
exactly the solutions
\begin{equation}
  \beta_a=\CovC^+\CovR{a}+z_a,
  \qquad z_a\in\kernel(\CovC),
  \label{eq:general-solution}
\end{equation}
where $\CovC^+$ is the Moore--Penrose inverse
(\cref{subsec:normal-equations}). Substituting into
\eqref{eq:target-coefficients} separates the identified component of the
target effect from uncertainty along these directions:
\begin{equation}
  \psi_T
  =\underbrace{b_T+\ellvec^\top \CovC^+\CovRdiff}_{\text{identified}}
  +\underbrace{\ellvec^\top(z_1-z_0)}_{\text{external-validity uncertainty}}.
  \label{eq:target-null-decomposition}
\end{equation}
\begin{corollary}[Target point identification]
\label{cor:point-id}
\label{cor:point-identification}
Assume unrestricted support for the potential outcomes.
Without additional restrictions, $\psi_T$ is point identified if and only if
\begin{equation}
  \ellvec\in\range(\CovC),
  \label{eq:point-id-condition}
\end{equation}
or, equivalently, $\ellvec^\top z=0$ for every $z\in\kernel(\CovC)$. If the
condition fails, the unrestricted target effect is unbounded from above and
below.
\textnormal{\hyperref[proof:point-id]{Proof in Appendix~\ref*{subsec:normal-equations}.}}
\end{corollary}

The span condition \eqref{eq:point-id-condition} is not to be confused with
overlap over unit-level covariates. A target can have excellent overlap in
unit-level covariates and still load on a direction of site-level
variables that no source
site identifies.  Conversely, a target site outside the convex hull
of the observed sites can still have a point-identified effect because
identification depends on the span of the residualized source
profiles, as in the no-$X$ example of \cref{fig:two-sites}. This population
singularity also affects estimation. Regressing each coordinate of the
site-level variables $C$ separately on unit-level covariates $X$ can introduce
residual components outside the source-site span, restoring full rank in
the sample and erasing the population indeterminacy. \Cref{sec:estimation}
estimates $\E_P[C\mid X]$ jointly so that the sample program retains the
population rank.

\subsection{Partial identification bounds after covariate adjustment}

We use the same restriction set $\mathcal R$ on
$\beta=(\beta_0^\top,\beta_1^\top)^\top$ as in \cref{subsec:examples}.
Replacing the no-$X$ source equations by the residualized normal equations
and adjusting the target objective gives
\begin{equation}
  \mathcal F
  =\left\{\beta:\ \CovC\beta_a=\CovR{a},\ a\in\{0,1\},
  \ \beta\in\mathcal R\right\},
  \label{eq:feasible-set}
\end{equation}
and we define the sensitivity interval by the extreme target effects
\begin{equation}
\begin{split}
  v_-&=b_T+\inf_{\beta\in\mathcal F}
  \ellvec^\top(\beta_1-\beta_0),\\
  v_+&=b_T+\sup_{\beta\in\mathcal F}
  \ellvec^\top(\beta_1-\beta_0).
\end{split}
\label{eq:endpoint-programs}
\end{equation}
Every $\beta\in\mathcal F$ reproduces the source conditional means.
The two optimization problems project this restricted identified set onto the target
effect, and we refer to $v_-$ and $v_+$ as the
\emph{partial identification bounds}. If $\mathcal F$ is convex, the partial identification bounds are finite, and the
outcome support is unrestricted, every value in $[v_-,v_+]$ is
attained by a model consistent with the data and the restriction. The interval
is then sharp (\cref{prop:mean-sharpness}). Support restrictions for bounded
or binary outcomes can shrink the attainable set. A finite union of polyhedra
can instead project to several disjoint intervals. In that case,
$[v_-,v_+]$ is their convex hull, and the separate intervals
should be retained to display the gaps.

Thus the restrictions retain their interpretation as limits on
site-level effects, while adjustment for $X$ changes both the source constraints
and the target objective. With no $X$, these programs reduce to
\eqref{eq:no-x-endpoint-programs}.

\section{Estimating partial identification bounds}
\label{sec:estimation}

Each partial identification bound in \eqref{eq:endpoint-programs} is the
value of a constrained optimization problem. We must estimate both the target-effect objective and
the linear equations that determine which coefficients fit the source data.
Estimating these equations requires care: small errors can make them appear
to determine coefficients that the population equations leave unrestricted.
The estimated partial identification bounds can then collapse to a point,
or the equations can become incompatible with the parameter restrictions. We therefore construct
estimates that preserve the linear dependencies implied by the
source site-level variables (\cref{subsec:rank-preservation}).

Once this structure is preserved, we address bias in the estimated objective
and constraints caused by estimating nuisance functions such as conditional
means. Flexible estimates of these functions can converge more slowly than
$n^{-1/2}$, so we use standard debiasing corrections to remove the first-order
effect of their estimation errors. Unlike the ATE, which is linear in the
outcome regressions for a fixed covariate distribution, our partial
identification bounds depend nonlinearly on nuisance functions through the
objective and constraints of an optimization problem. A natural approach is
to solve the optimization problem in each fold and then apply a one-step
bias correction. But nuisance error can make even
one fold's constraints infeasible, leaving no estimate to correct. Our
approach reverses this order: we debias and pool the inputs, then solve a
single optimization problem. Feasibility is
therefore required only at the pooled inputs (\cref{subsec:two-estimators}).

Sampling error also changes which coefficients satisfy the estimated
constraints, so standard errors must account for how these changes affect
the optimized value. We construct standard errors that remain well defined
despite nonunique dual variables (\cref{subsec:crossfitting}). The simulations
in \cref{sec:simulation} show why pooling before optimization matters in
small samples: it avoids the disagreement across foldwise solutions that
can prevent one-step inference.

\subsection{Preserving the rank of the equality system}
\label{subsec:rank-preservation}

Both sides of the linear equations \eqref{eq:profiled-moments} are
estimated using residuals from regressing site-level variables $C$ on
unit-level covariates $X$. The fitted conditional mean
$\what m_C(x)$ estimates $m_C(x)=\E_P[C\mid X=x]$ and must respect the linear
relationships among the source site-level variables. If we fit the coordinates
of $m_C$ separately, the
vector $\what m_C(X_i)$ need not remain in the affine span of the source-site
profiles. The residuals $C_i-\what m_C(X_i)$ can then acquire components in
directions of site-level variables that do not vary across the observed sites. Even small
components in these directions can make $\whatCovC$ full rank. The plug-in
equalities may consequently replace an affine population feasible set with a
single coefficient vector. If that vector violates the restrictions in
$\mathcal R$, the sample program becomes infeasible. Regression error can
therefore create spurious point identification or infeasibility. Keeping
the fitted residuals in the source-site span prevents these errors.

The finite set of source site-level variables gives a simple way to preserve rank. Since
$C$ takes one of the values $c_1,\ldots,c_S$, its conditional mean $m_C(X)$ is
a weighted average of these same site profiles. Subtracting the conditional
mean therefore removes their common location and leaves a vector in
\begin{equation}
  \mathcal V_C=\operatorname{span}\{c_s-c_1:s=2,\ldots,S\},
  \label{eq:source-context-span}
\end{equation}
so $C-m_C(X)\in\mathcal V_C$. It follows that
$\range(\CovC)\subseteq\mathcal V_C$ and $\CovR{a}\in\mathcal V_C$. The
population equality system thus varies only along the directions of
site-level variables present in the source sites.

We impose the same restriction on the fitted residuals. We estimate the full
vector of site probabilities $p_s(x)=\pr(C=c_s\mid X=x)$ with a joint
classifier and set
\begin{equation}
  \what m_C(x)=\sum_{s=1}^S c_s\what p_s(x).
  \label{eq:context-mixture-regression}
\end{equation}
The fitted probabilities sum to one, so both $C_i$ and
$\what m_C(X_i)$ are affine combinations of the source profiles. Their
difference lies in $\mathcal V_C$ exactly, regardless of the classifier's
estimation error. We use this same residual in every estimated moment involving
$C$. Estimation error can then move these moments within $\mathcal V_C$, but
it cannot introduce a direction orthogonal to the observed site variation.

To remove the redundant equations, let $r=\dim(\mathcal V_C)$ and let
$B_C\in\R^{p\times r}$ have orthonormal columns spanning $\mathcal V_C$. The
sample moments can be written as
\[
  \whatCovC=B_C\what\Omega B_C^\top,
  \qquad
  \whatCovR{a}=B_C\what\varsigma_a,
\]
where $\what\Omega=B_C^\top\whatCovC B_C$ and
$\what\varsigma_a=B_C^\top\whatCovR{a}$. We give the solver the reduced equalities
\[
  \what\Omega B_C^\top\beta_a=\what\varsigma_a,
  \qquad a\in\{0,1\}.
\]
Under the within-span eigenvalue condition in \cref{ass:sampling},
$\what\Omega$ is invertible with probability approaching one. On this event,
the right-hand side is in the range of $\whatCovC$, and the reduced system has
full row rank. The reduced equalities remove only the redundant rows of the
sample analogue of \eqref{eq:profiled-moments}. They therefore define the same feasible
coefficient vectors and the same partial identification bound while preserving the population rank.

\subsection{Debiasing the inputs before optimization}
\label{subsec:two-estimators}

Preserving rank does not ensure that each fold's estimated source equations
are compatible with the parameter restrictions. The natural one-step
approach, which optimizes in each fold before correcting nuisance bias,
can therefore fail before the correction is even defined.

We avoid this requirement by correcting the objective and source moments
directly using regression residuals. We then pool the corrected inputs
across folds and optimize once, so only the pooled optimization problem
must be feasible. The construction below first gives the corrections for
the source and target inputs, then combines them into the estimator.
The optimizer and dual variables enter only the standard error, for which
our theory requires consistency (\cref{sec:asymptotics}). We give the
one-step benchmark in \cref{app:one-step-benchmark} and examine the practical
consequences of optimizing before pooling in \cref{sec:simulation}.

Write $v(\Gamma)$ for the lower partial identification bound in \eqref{eq:endpoint-programs}, where
\begin{equation}
  \Gamma=(b_T,\ellvec,\CovC,\CovR{0},\CovR{1}).
  \label{eq:program-input-vector}
\end{equation}
The upper partial identification bound uses the corresponding maximization. For a finite union of
polyhedra, we optimize over each polyhedron and take the outer extrema. Our scores use the
joint and marginal outcome regressions
\begin{equation}
  \mu_a(x,c)=\E_P[Y\mid A=a,X=x,C=c],
  \qquad \mu_a(x)=\E_P[R_a\mid X=x],
  \label{eq:nuisance-definitions}
\end{equation}
along with $m_C$. We estimate the inputs with orthogonal scores, whose
expectations are insensitive to first-order nuisance error, and evaluate
each score on observations held out from fitting
\citep{bickel1993efficient,newey1990semiparametric,tsiatis2006semiparametric,chernozhukov2018double}.

\paragraph{Debiased program inputs.}
\label{par:debiased-inputs}
Randomized treatment assignment makes the pseudo-outcome $R_a$ noisy.
We use the standard augmented outcome based on the joint outcome
regression $\mu_a(X,C)$:
\begin{equation}
  U_a=\mu_a(X,C)+\frac{\one\{A=a\}}{e_a(X,C)}\{Y-\mu_a(X,C)\}.
  \label{eq:augmented-outcome}
\end{equation}
Both $R_a$ and $U_a$ have conditional mean $\mu_a(X,C)$, so they define the
same population moments. We use $U_a$ to reduce variation from randomized
treatment assignment. The treatment probabilities are known from the
randomized design.

Let $O_1,\ldots,O_n\simiid P$ be the source observations and let
$X_{T,1},\ldots,X_{T,n_T}\simiid Q_T$ be an independent covariate sample from the
target. We fit the joint and marginal outcome regressions separately.
Suppressing fold indices, the source moments are
\begin{equation}
\begin{split}
  \what U_{a,i}&=\what\mu_a(X_i,C_i)
  +\frac{\one\{A_i=a\}}{e_a(X_i,C_i)}
  \{Y_i-\what\mu_a(X_i,C_i)\},\\
  \whatCovC&=\frac{1}{n}\sum_{i=1}^n \{C_i-\what m_C(X_i)\}\{C_i-\what m_C(X_i)\}^\top,\\
  \whatCovR{a}&=\frac{1}{n}\sum_{i=1}^n
  \{C_i-\what m_C(X_i)\}\{\what U_{a,i}-\what\mu_a(X_i)\}, \qquad a\in\{0,1\}.
\end{split}
\label{eq:orthogonal-source-moments}
\end{equation}
Iterated expectations set
the first-order derivatives with respect to $m_C$ and the marginal $\mu_a$ to zero.
Randomization sets the derivative with respect to the joint $\mu_a$ to zero because
$\E[\one\{A=a\}/e_a(X,C)\mid X,C]=1$.

The target inputs average regressions under $Q_T$, but their residual
corrections use source observations. We reweight those residuals using the
density ratio
\begin{equation}
  w(x)=\frac{dQ_T}{dP_X}(x).
  \label{eq:density-ratio}
\end{equation}
Each augmented transport score combines a target-sample regression average
with a reweighted source-sample residual correction:
\begin{equation}
\begin{split}
  \what b_T
  &=\frac{1}{n_T}\sum_{i=1}^{n_T}
    \{\what\mu_1(X_{T,i})-\what\mu_0(X_{T,i})\}\\
  &\quad+\frac{1}{n}\sum_{i=1}^n \what w(X_i)
  \big[\{\what U_{1,i}-\what\mu_1(X_i)\}
  -\{\what U_{0,i}-\what\mu_0(X_i)\}\big],\\
  \what\ell_T
  &=c_T-\frac{1}{n_T}\sum_{i=1}^{n_T}\what m_C(X_{T,i})
  -\frac{1}{n}\sum_{i=1}^n\what w(X_i)\{C_i-\what m_C(X_i)\}.
\end{split}
\label{eq:orthogonal-target-moments}
\end{equation}
The density-ratio error enters the bias only through its product with
regression error. Together with the source scores, this leaves a remainder
that is $o_p(n^{-1/2})$ under the product-rate conditions in
\cref{ass:nuisance-plugin}.\footnote{If the target law $Q_T$ is known exactly, say from a census,
we replace each empirical target average in
\eqref{eq:orthogonal-target-moments} by the exact $Q_T$ integral of the
fitted regression. The source corrections are unchanged, and no target
sampling uncertainty enters the theory.}

Identification requires $Q_T\ll P_X$; inference also requires the
density-ratio moment conditions in \cref{sec:asymptotics}. Neither condition
requires the target site $c_T$ to lie in the source support.

\paragraph{Pooling the corrected inputs.}
Collect the nuisance functions as
\begin{equation}
  \eta=(m_C,\mu_0,\mu_1,w),
  \label{eq:nuisance-vector}
\end{equation}
where each outcome regression enters through its joint and marginal forms.
We split the source and target samples into paired folds, fit the nuisance
functions outside each fold, and evaluate
\eqref{eq:orthogonal-source-moments} and \eqref{eq:orthogonal-target-moments} on
its held-out observations. Pooling these scores by observation gives
$\what\Gamma$. We substitute these inputs into
\eqref{eq:endpoint-programs}, so the estimator of the lower partial identification bound is
\begin{equation}
  \vhatE{\Eone}{-}=v(\what\Gamma).
  \label{eq:empirical-endpoint-programs}
\end{equation}
We retain the optimizer and dual variables for the standard error. The
point estimate requires no further correction.
\Cref{alg:crossfit-plugin} summarizes the estimator.

\begin{algorithm}[t]
\caption{Proposed estimator of partial identification bounds}
\label{alg:crossfit-plugin}
\begin{algorithmic}[1]
\Require source sample $\{O_i\}_{i=1}^n$, target covariates
$\{X_{T,i}\}_{i=1}^{n_T}$ (or a known $Q_T$), target site-level variable $c_T$, basis $B_C$ for
$\mathcal V_C$, restriction set $\mathcal R$, fold count $K$
\State Split both samples into $K$ paired balanced folds.
\For{$k=1,\ldots,K$}
  \State Fit
  $\what\eta_k$ as in \eqref{eq:nuisance-vector} on data
  outside the $k$th folds, fitting joint site probabilities so that
  $C-\what m_{C,k}(X)\in\mathcal V_C$.
  \State Evaluate the debiased scores in
  \eqref{eq:orthogonal-source-moments} and
  \eqref{eq:orthogonal-target-moments} on the held-out observations.
\EndFor
\State Pool the scores by observation to form the estimate
$\what\Gamma=(\what b_T,\what\ell_T,\whatCovC,\whatCovR{0},\whatCovR{1})$.
\State Express each arm's equality constraints as
$B_C^\top\whatCovC\beta_a=B_C^\top\whatCovR{a}$, keeping $\beta_a\in\R^p$.
\State Solve the lower and upper programs in \eqref{eq:endpoint-programs}
at $\what\Gamma$ to obtain $\vhatE{\Eone}{-}$ and $\vhatE{\Eone}{+}$.
\State Record each value, optimizer, reduced equality dual variable, and solver diagnostics.
\State Report $[\vhatE{\Eone}{-},\vhatE{\Eone}{+}]$ and Wald intervals built from the empirical
variance of \eqref{eq:source-if} and \eqref{eq:target-if}, using each partial identification bound's
own optimizer and dual variable.
\end{algorithmic}
\end{algorithm}

\FloatBarrier

\subsection{Standard errors}
\label{subsec:crossfitting}

Sampling error changes both the target-effect objective and the set of
coefficients satisfying the source constraints. Standard errors must
therefore account for how an estimated constraint shifts the optimal value,
even after we have removed first-order nuisance bias from the inputs.
The optimizer measures sensitivity to the objective, while the equality
dual variables measure sensitivity to the constraints.

This creates an apparent ambiguity: the redundant source equations admit
multiple dual solutions. We resolve it using the same reduced equality
system that preserves rank. All admissible perturbations of the source equations
lie in $\mathcal V_C$, so only a dual variable's projection onto this span
can affect the partial identification bound. The dual $\bar\lambda_a$ from
the reduced system in \cref{subsec:rank-preservation} gives the representative
\begin{equation}
  \lambda_a=B_C\bar\lambda_a.
  \label{eq:min-norm-multiplier}
\end{equation}
This is the minimum-norm full dual and has a unique population target under
our regularity conditions. Thus the standard error does not depend on an
arbitrary choice among full dual solutions.

Let $\beta^*=(\beta_0^*,\beta_1^*)$ be the population optimizer for the lower
partial identification bound. An envelope argument combines the objective
and constraint sensitivities to give
\begin{equation}
\begin{gathered}
  \what v-v
  =(\what b_T-b_T)+(\what\ell_T-\ellvec)^\top(\beta^*_1-\beta^*_0)\\
  +\sum_{a\in\{0,1\}}\lambda_a^\top
  \bigl\{(\whatCovR{a}-\CovR{a})-(\whatCovC-\CovC)\beta^*_a\bigr\}
  +o_p(n^{-1/2}).
\end{gathered}
  \label{eq:envelope-heuristic}
\end{equation}
Substituting the input scores gives the source influence function
\begin{equation}
\begin{split}
  \phi(O)
  &=w(X)\Big[\{U_1-\mu_1(X)\}-\{U_0-\mu_0(X)\}
  +(\beta^*_1-\beta^*_0)^\top\{m_C(X)-C\}\Big]\\
  &\quad+\lambda_0^\top(C-m_C(X))
  \bigl\{U_0-\mu_0(X)-(C-m_C(X))^\top\beta^*_0\bigr\}\\
  &\quad+\lambda_1^\top(C-m_C(X))
  \bigl\{U_1-\mu_1(X)-(C-m_C(X))^\top\beta^*_1\bigr\},
\end{split}
\label{eq:source-if}
\end{equation}
and, when the target covariates are sampled,
\begin{equation}
  \phi_T(X_T)
  =\mu_1(X_T)-\mu_0(X_T)
  +\{c_T-m_C(X_T)\}^\top(\beta^*_1-\beta^*_0)-v.
  \label{eq:target-if}
\end{equation}
Both functions have mean zero. The first line of \eqref{eq:source-if}
accounts for estimation of the target inputs, and the remaining lines
account for the source constraints. Equation \eqref{eq:target-if} captures
sampling variation in the target covariate distribution.

For the upper partial identification bound, we use its own optimizer and
dual; a dual obtained by minimizing the negative objective must be
multiplied by $-1$. \Cref{sec:asymptotics} establishes the expansion and
consistency of these optimization estimates on the domain of moments
generated by the observed sites.

We evaluate the influence functions at the cross-fitted nuisances and each
partial identification bound's optimizer and dual. If $\what\sigma_j^2$ and
$\what\sigma_{T,j}^2$ are their empirical source and target variances, the
standard error for endpoint $j\in\{-,+\}$ is
$\what{\mathrm{se}}_j=(\what\sigma_j^2/n+
\what\sigma_{T,j}^2/n_T)^{1/2}$. Its pointwise Wald interval is
$\what v_j\pm z_{1-\alpha/2}\what{\mathrm{se}}_j$.
When $Q_T$ is known and integrated exactly, we omit the target variance term.

\subsection{Numerical evidence for pooling before optimization}
\label{sec:simulation}
\label{sec:simulations}

We study a finite-sample obstacle to the one-step correction: \emph{sign
mismatch} across folds. Under effect hierarchy \eqref{eq:effect-ordering},
each sign pattern of the direct coefficients defines a linear program.
A sign mismatch occurs when the folds select different optimizing sign
patterns for either the lower or upper partial identification bound. The
standard one-step correction (\cref{alg:crossfit-onestep}) requires agreement across
folds and withholds an interval when this condition fails, even if every
fold's optimization problem is feasible. Our estimator
(\cref{alg:crossfit-plugin}) pools the corrected inputs before optimizing,
so it eliminates this source of failure. At $n=500$, sign mismatch prevents
one-step inference in 89 percent of replications, while our approach permits
inference in every replication. We examine how this mismatch and interval
coverage change with sample size.

We focus on the design most likely to expose this problem: partial
identification with nonlinear nuisance functions. Outcomes follow
$Y=g_A(X)+C^\top\beta_A+\varepsilon$, where $\varepsilon\sim N(0,1)$, and
treatment probabilities vary across sites from $0.35$ to $0.65$. The
unit-level covariates include a binary, a binomial, and a Gaussian
variable. Their relationships with site and outcome are nonlinear, and we
estimate the nuisance functions with gradient-boosted trees. Each source
sample of size $n$ is paired with an independent target covariate sample of
the same size. Four sites span only three of six site-level terms, so three
directions remain unidentified. We obtain finite partial identification bounds by imposing the box
$|\beta_{a,j}|\leq2.5$ and effect hierarchy~\eqref{eq:effect-ordering} at
$\kappa=1$, which requires each higher-order site-level coefficient to be no larger in
absolute value than every direct site-level coefficient in the same treatment
arm. Conditioning on the signs of the three direct coefficients in each arm
gives 64 linear programs. We compare against the benchmark using the same data draw, outer
folds, nuisance learner class, restrictions, and solver. We use 100 replications
at each $n\in\{500,1{,}000,2{,}500,5{,}000,10{,}000\}$.

\begin{figure}[t]
  \centering
  \includegraphics[width=0.88\textwidth]{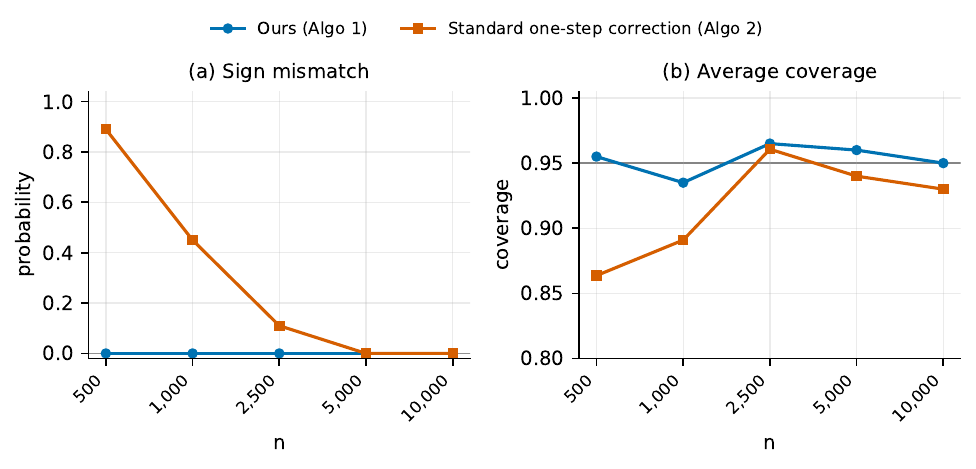}
  \caption{Sign mismatch and coverage with boosted nuisance fits. We compare our
    estimator (\cref{alg:crossfit-plugin}) with the standard one-step
    correction (\cref{alg:crossfit-onestep}).
    Panel~(a) reports the probability of sign mismatch for either partial
    identification bound. The standard one-step correction withholds an interval
    when its four outer folds disagree on the optimizing coefficient signs;
    our estimator eliminates this requirement by pooling before
    optimization. Every fold-level optimization problem in these simulations
    is feasible. Panel~(b) reports pointwise coverage averaged across the
    lower and upper partial identification bounds; coverage for
    the standard one-step correction is conditional on no sign mismatch.
    The gray lines mark zero mismatch and 95 percent coverage. Sign mismatch
    often prevents one-step inference in small samples. At larger sample
    sizes, mismatch disappears and the procedures have similar coverage.}
  \label{fig:sim-matched-boosted}
\end{figure}

Panel~(a) of \cref{fig:sim-matched-boosted} shows that the probability
of sign mismatch for the standard one-step correction is 0.89 at $n=500$, 0.45
at $n=1{,}000$, and 0.11 at $n=2{,}500$; no mismatch occurs from $n=5{,}000$
onward. Every fold-level optimization problem is feasible, so the failures
arise entirely from sign mismatch. Panel~(b) shows poor conditional coverage
for the standard one-step correction in the two smallest samples, based on the
11 and 55 replications without sign mismatch. From $n=5{,}000$ onward,
both algorithms permit inference in every replication and their coverage
is similar. These results favor our estimator because pooling
before optimization eliminates sign mismatch across folds. The demonstrated
advantage concerns whether inference is possible in small samples; coverage
is similar once the folds agree, and this design does not compare the
methods under a single polyhedron.

\FloatBarrier

\section{Asymptotic theory}
\label{sec:asymptotics}

We develop debiased inference for partial identification bounds defined by
optimization problems with estimated equality constraints. Estimation
changes both the objective and which coefficients satisfy the source
equations. Even with the objective held fixed, errors in the constraints
shift the feasible set and hence the optimized value. The equality dual
variables measure this sensitivity, so they must enter the influence
function and standard error. Here the equality system is also rank
deficient: arbitrary perturbations can destroy feasibility, and redundant
equations make the full dual variables nonunique.

To our knowledge, we provide the first debiased Gaussian inference result
for partial identification bounds whose rank-deficient equality system has
both an estimated coefficient matrix and an estimated right-hand side,
allowing nonunique full equality dual variables.\footnote{Related results
address different constraint structures. \citet{benmichael2025partial}
develops debiased inference for conditional linear programs with a known
equality-constraint matrix. \citet{kim2022counterfactual} study inference
for optimization solutions with estimated constraints under linear
independence of the active constraint gradients.}
We first establish differentiability along the moment perturbations
preserved by our estimator and show that all full dual solutions give the
same derivative on this domain. We then prove consistency of the optimizer
and reduced dual estimates and combine the derivative with orthogonal
scores for both the objective and constraints. This gives valid inference
while requiring only consistency of the optimization estimates.

The first step is to control how the estimated equalities change the
feasible set. We need to differentiate only along directions in which the
estimated moments can vary. An arbitrary perturbation of \eqref{eq:profiled-moments}
can move $\CovR{a}$ outside $\range(\CovC)$ and make the equality infeasible.
Our estimator preserves the source-site span $\mathcal V_C$,
so we restrict the tangent space to perturbations that preserve this span.
Using the basis $B_C$ from \cref{subsec:rank-preservation}, this domain is
\begin{equation}
  \mathcal G_C
  =\bigl\{(b,\ell,B_C\Omega B_C^\top,B_C\varsigma_0,B_C\varsigma_1):
    (b,\ell,\Omega,\varsigma_0,\varsigma_1)\in\mathcal U\bigr\},
  \label{eq:site-moment-domain}
\end{equation}
where $\mathcal U=\R\times\R^p\times\mathbb S_{++}^r\times\R^r\times\R^r$
is open in the reduced coordinates
$\gamma=(b,\ell,\Omega,\varsigma_0,\varsigma_1)$. Positive definiteness is required
only within $\mathcal V_C$; $\CovC$ may remain singular.

The dual variables translate these changes in the equalities into changes
in the optimized value. Their nonuniqueness would leave the influence
function ambiguous if different dual solutions assigned different weights
to the same estimation error. Restricting the perturbations resolves this
problem. Adding a vector in $\kernel(\CovC)$ leaves the optimality
conditions unchanged, but this component is orthogonal to every admissible
perturbation of the source equations. Only the projection of the dual onto
$\mathcal V_C$ affects sensitivity to the constraints. We therefore use
$\lambda_a=B_C\bar\lambda_a$, where $\bar\lambda_a$ is the reduced equality
dual, and establish convergence to this minimum-norm representative.

For one polyhedron, write $\mathcal R=\{\beta:G\beta\leq h\}$. The following
assumption ensures that the perturbed equalities remain compatible with
these restrictions locally and that the optimizing coefficients remain
bounded. It also supplies a relative interior point. Uniqueness is required for
the population optimizer and the reduced dual, so perturbations select a
single linear derivative. Without this uniqueness, the partial identification
bound can have only directional derivatives, and Gaussian inference need
not apply.

\begin{assumption}[Regular partial identification bound]
  \label{ass:regular-endpoint}
  For each partial identification bound and its population-optimal polyhedron: (i) throughout a
  neighborhood $\mathcal U_0\subset\mathcal U$ of the population reduced input
  $\gamma^*$, the program is feasible, its optimal value is attained and finite, and is locally bounded as a function of the moments; (ii) at
  $\gamma^*$, relative Slater holds after redundant equality rows are removed
  and after deleting only inequalities that are implied by the reduced
  equalities throughout $\mathcal U_0$; (iii) the population lower level set
  through a Slater point is bounded after orienting the partial identification bound as a
  minimization; and (iv) the population optimizer and the reduced KKT dual
  pair are unique at $\gamma^*$. The restriction matrices $(G,h)$ are held
  fixed along every moment perturbation.
\end{assumption}

Under \cref{ass:regular-endpoint}, the partial identification bound $v$
is Hadamard differentiable on $\mathcal G_C$ tangentially to its tangent
space. For an admissible perturbation
$\dot\Gamma=(\dot b,\dot\ell,\dot\Sigma,\dot d_0,\dot d_1)$,
\begin{equation}
  Dv(\Gamma)[\dot\Gamma]
  =\dot b+\dot\ell^\top(\beta^*_1-\beta^*_0)
  +\sum_{a\in\{0,1\}}\lambda_a^\top(\dot d_a-\dot\Sigma\beta^*_a),
  \label{eq:site-domain-derivative}
\end{equation}
where $\beta^*$ is the population optimizer. The first two terms capture
error in the objective. The residual $\dot d_a-\dot\Sigma\beta_a^*$ measures
how the perturbed equality fails at the population optimizer, and its dual
weight gives the resulting first-order change in the partial identification
bound. Thus the derivative accounts for estimation error on both sides of
the equality. Since this residual lies in $\mathcal V_C$, every full dual
representative gives the same derivative. \Cref{prop:value-derivative}
proves this formula, allowing us to apply the tangential delta method
\citep[Theorem~20.8]{van2000asymptotic} despite the singular equality system.

The same regularity conditions ensure consistency of the optimizer and
reduced dual estimates as the input estimates converge
(\cref{lem:optimizer-stability,prop:kkt-consistency}), using
perturbation arguments from
\citet{bonnans2000perturbation,bonnans1998guided,rockafellar1998variational}.
The empirical checks appear in \cref{subsec:implementation-diagnostics}.
For our estimator, these optimization estimates enter only the
standard error; they do not enter a separate correction to the point
estimate. Their consistency is therefore sufficient, with no additional
rate requirement.

The derivative determines how uncertainty in the estimated equalities
combines with uncertainty in the objective. The source influence function
combines both contributions, so its variance retains their covariance. An
independent target sample contributes a separate variance term. We impose
the following sampling conditions for these joint expansions.

\begin{assumption}[Sampling and overlap]
  \label{ass:sampling}
  The source observations are i.i.d. from $P$. In the sampled-target design, the
  target covariates are i.i.d. from $Q_T$ and independent of the source
  sample. In the known-law design, $Q_T$ is fixed and target integrals are
  evaluated exactly. Treatment probabilities are known and bounded away from
  zero. The density ratio $w=dQ_T/dP_X$ exists. The source sites have fixed
  support $\{c_1,\ldots,c_S\}$, each relevant site probability is positive, and
  the covariance of $B_C^\top\{C-m_C(X)\}$ has its smallest eigenvalue bounded
  away from zero. For some $q>4$, the pseudo-outcomes, site-level basis, density
  ratio, and influence functions in \eqref{eq:source-if} and
  \eqref{eq:target-if} have finite $q$th moments.
\end{assumption}

The eigenvalue condition keeps the source moments at the required rank,
while the $q>4$ moment condition controls the nuisance remainders and permits
a two-sample central limit theorem.

Debiasing must remove first-order nuisance error from the equality
constraints as well as the objective. Once their scores are orthogonal,
applying the derivative leaves products of nuisance estimation errors as
the remaining bias. Cross-fitting evaluates each score on
observations independent of the nuisance estimates, avoiding Donsker
conditions on the fitted function classes. The next assumption makes these
products negligible relative to $n^{-1/2}$; it imposes no comparable rate
on the optimizer or dual estimates.

\begin{assumption}[Nuisance rates for plug-in debiasing]
  \label{ass:nuisance-plugin}
  Conditional on each training sample of the cross-fitting scheme in
  \cref{subsec:two-estimators}, the true and estimated nuisance functions are
  uniformly bounded in $L_q(P)$ and the cross-fitted estimates are consistent in
  $L_2(P)$. In addition, the marginal $\what\mu_a(X)$ and $\what m_C$ are
  consistent in $L_2(Q_T)$ and uniformly bounded in $L_q(Q_T)$. Norms of $\mu_a$
  below refer to its marginal form. Norms below are under $P$ unless marked
  otherwise. The source errors satisfy
  \begin{equation}
    \begin{split}
      &\|\what w-w\|_2
      \left(\|\what\mu_0-\mu_0\|_2+\|\what\mu_1-\mu_1\|_2
        +\|\what m_C-m_C\|_2\right)=o_p(n^{-1/2}),\\
      &\|\what m_C-m_C\|_2^2=o_p(n^{-1/2}),\qquad
      \|\what m_C-m_C\|_2\|\what\mu_a-\mu_a\|_2=o_p(n^{-1/2})
    \end{split}
    \label{eq:product-rates}
  \end{equation}
  for $a\in\{0,1\}$. It is sufficient that $\what m_C$, the marginal
  $\what\mu_0,\what\mu_1$, and $\what w$ each converge at rate
  $o_p(n^{-1/4})$. The joint outcome regressions $\what\mu_a(X,C)$ need only be
  consistent, and no rate is imposed on the estimated optimizer or dual variable
  beyond $o_p(1)$.
\end{assumption}

Applying \eqref{eq:site-domain-derivative} to the cross-fitted input
expansions gives the source and target influence functions in
\eqref{eq:source-if} and \eqref{eq:target-if}. In particular, the dual-weighted
source terms account for the movement of the estimated equality constraints;
omitting them would generally give an incorrect asymptotic variance. When the restriction set is
a finite union of polyhedra, \cref{ass:branch-separation} ensures that the polyhedron
attaining each partial identification bound is locally unchanged
(\cref{prop:value-derivative}), so the same expansion applies. For sampled
target covariates, let $n_T/n\to\zeta\in(0,\infty]$; $\zeta=\infty$ covers
a target sample large relative to the source sample.

\begin{theorem}[Inference on the partial identification bounds]
  \label{thm:endpoint-clt}
  \label{thm:plugin-clt}
  \label{thm:setup-master}
  Suppose \cref{ass:model,ass:sampling,ass:regular-endpoint,ass:nuisance-plugin}
  hold and let $j\in\{-,+\}$ index the two partial identification bounds. For a polyhedral restriction,
  our estimator $\vhatE{\Eone}{j}$ (\cref{alg:crossfit-plugin}) satisfies
  \begin{equation}
    \sqrt n(\what v_j-v_j)
    =\frac{1}{\sqrt n}\sum_{i=1}^n\phi_j(O_i)
    +\frac{\sqrt n}{n_T}\sum_{i=1}^{n_T}\phi_{T,j}(X_{T,i})+o_p(1)
    \rightsquigarrow N(0,V_j).
    \label{eq:asymptotic-linearity}
  \end{equation}
  Here
  \begin{equation}
    V_j=\var_P\{\phi_j(O)\}
    +\zeta^{-1}\var_{Q_T}\{\phi_{T,j}(X_T)\}.
    \label{eq:setup-master-variance}
  \end{equation}
  Its sample analogue is consistent. Under \cref{ass:branch-separation}, the same holds for the finite-union
  partial identification bound. The upper partial identification bound uses its own optimizer
  and dual variable in \eqref{eq:source-if} and \eqref{eq:target-if}. If $Q_T$
  is known and each fold-specific fitted regression is integrated exactly
  under $Q_T$, the same conclusions hold after deleting the target sum in
  \eqref{eq:asymptotic-linearity} and the target component of $V_j$.
\textnormal{\hyperref[subsec:proof-theorem-three]{Proof in Appendix~\ref*{subsec:proof-theorem-three}.}}
\end{theorem}

The theorem establishes Gaussian inference while accounting for sampling
error in the estimated feasible set. The Wald intervals in
\cref{subsec:crossfitting} incorporate the contribution of the equality
constraints through consistently estimated dual weights, without requiring
a unique full dual solution or a convergence rate for the optimization
estimates beyond consistency. For the partial identification bound functional of
the original moments, $\phi_j$ and $\phi_{T,j}$ are the canonical gradients
when the observed source law is unrestricted apart from known randomization
probabilities and fixed support of site-level variables, and the target covariate law is
unrestricted. The variance in \cref{thm:endpoint-clt} is therefore the minimum
asymptotic variance among regular estimators in this observed-data model.

\section{Case study: microcredit across countries}
\label{sec:case-study}

We now return to the seven microcredit experiments of
\citet{meager2019understanding}. To illustrate our method we select one of the seven countries as a
target site whose program characteristics and household covariates are observed but whose
experimental outcomes are not. The other six experiments serve as source
sites. We estimate the target effect from them and consult the held-out
experiment only afterward. This leave-one-country-out exercise asks what a
policymaker could have concluded before the target experiment reported its
results.

We illustrate our framework on Bosnia and Morocco. These two sites have
similar ground truth effects---positive effects on household revenues: $79.8$
revenue units in Bosnia and $80.9$ in Morocco---yet vary substantially in
terms of their similarity to the rest of the sites. We shall see how our
sensitivity approach can show whether a conclusion's external validity is
anchored in observed cross-site variation or in extrapolation assumptions, and
how sensitive it is to the choice of source sites.  In particular, our method
flags the inability to extrapolate in the case of Morocco, and provides some
initial evidence that data from other sites can inform the treatment effect in
Bosnia.

\subsection{Empirical design}
\label{subsec:spec}

We use household revenues on the cross-site consistent scale supplied with the
study archive. Unit-level covariates include age, gender, household size, and
indicators for no formal education, prior business ownership, and urban
residence. We standardize these variables within each site. The
site-level variables contain promotion type, annual percentage rate (APR), and average
loan size as a share of income, together with their squares and pairwise
interactions. We replace each continuous term with its rank among the six
source sites and center the ranks at the source site median. This
produces $p=9$ site-level coefficients. We use three-fold cross-fitting, XGBoost nuisance fits, and a
regularized multinomial site classifier for $m_C$. We aggregate
influence-function standard errors at each country's randomization unit.

The six source-site profiles provide only five independent directions
of site-level variables. The covariance matrix $\CovC$ has rank $5$ for both
targets, leaving exactly four combinations of the nine coefficients in
each treatment arm unidentified. To assess how far each target extends
beyond the observed site variation, we subtract the mean source-site
profile from its target profile. The component outside the source-site
span accounts for 89\% of this difference vector's length for Morocco and
46\% for Bosnia. This geometric diagnostic shows that Morocco requires
more extrapolation beyond the observed site variation. The parameter
restrictions determine how much the unidentified coefficient combinations
can change the target effect.

We consider three parameter restrictions. First, we impose the magnitude cap
on $\|\beta\|_\infty$ given by one standard deviation of the outcome. Each
basis column is scaled by its standard deviation across the source sites, so
the cap has the same interpretation across site-level coefficients. Second, we restrict the coefficients on the APR term and
the APR interactions in the treatment shift $\beta_1 -\beta_0$ to be
nonpositive, while leaving the coefficient on $\mathrm{APR}^2$ and the control arm unrestricted. At the median source-site profile, this says that
an increase in the cost of credit does not raise the benefit of
access to credit.
Third, we impose effect hierarchy at $\kappa=1$. This states that within each treatment arm, every higher-order
coefficient must be no larger in magnitude than every direct coefficient. This
restriction can sharply limit the interaction terms when one direct effect is
close to zero.

For each specification, we ask whether any coefficient vector satisfies
both the pooled estimated source equations and these parameter restrictions.
We call the optimization problem \emph{feasible} when such a vector exists
and \emph{infeasible} otherwise. Our estimator (\cref{alg:crossfit-plugin}) preserves the rank
of the source equations and pools before optimization, but it cannot ensure
that the pooled equations are compatible with the chosen restrictions.
Sampling and nuisance estimation errors can change this compatibility.

\begin{figure}[t]
  \centering
  \includegraphics[width=0.96\textwidth]{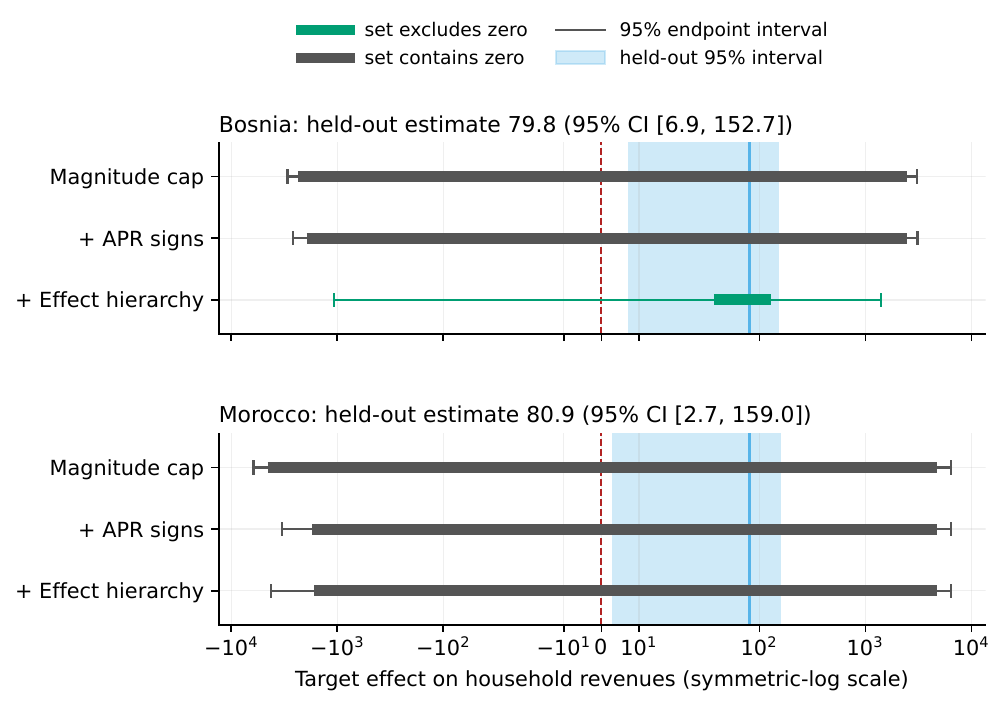}
  \caption{Successive restrictions for Bosnia and Morocco. We first
  impose the magnitude cap. We then add the APR sign restriction. We add
  effect hierarchy last. We use thick bars for the outer identified intervals
  and thin lines for pointwise 95\% intervals for the endpoints. The blue
  bands are the 95\% intervals from the held-out experiments. The APR sign restriction narrows
  both intervals but does not determine either sign. Effect hierarchy then
  moves the Bosnia interval above zero. The Morocco sign remains unresolved.}
  \label{fig:case-pair-bosnia-morocco}
\end{figure}

\subsection{What the source experiments support}
\label{subsec:case-results}

\cref{fig:case-pair-bosnia-morocco} follows the same restriction order for Bosnia and Morocco. 
The magnitude cap alone leaves both signs unresolved. The APR sign
restriction reduces the Bosnia width by 8.9\% and the Morocco width by
30.3\%. Neither interval has a fixed sign at this step. Effect hierarchy is the
decisive step for Bosnia. It reduces the remaining width by 97.9\%, from
$4389.5$ to $90.2$, and gives $[37.1,\,127.3]$. This interval is positive and
contains the held-out estimate of $79.8$. In Morocco, effect hierarchy reduces
the remaining width by only 1.1\% and gives $[-1653.1,\,4763.4]$. The APR
sign restriction therefore accounts for most of Morocco's limited reduction.
Effect hierarchy has the largest effect in Bosnia.

The positive estimated identified set for Bosnia does not establish a
positive effect with statistical confidence. After all three restrictions,
the outer limits of the pointwise 95\% intervals for its lower and upper
partial identification bounds are $[-1069.6,\,1407.0]$, which include zero.
The estimates also depend on how observations are partitioned for
cross-fitting. In four of seven partitions, no coefficient vector satisfies
both the pooled estimated source equations and the restrictions. In the
other three, the estimated identified set is positive, with lower partial
identification bounds from $0.3$ to $37.1$. These checks show that the
positive conclusion depends on the estimated equations being compatible
with the restrictions and remains subject to substantial sampling
uncertainty. Morocco's interval contains zero in the same checks. 

\begin{figure}[tbp]
  \centering
  \includegraphics[width=\textwidth]{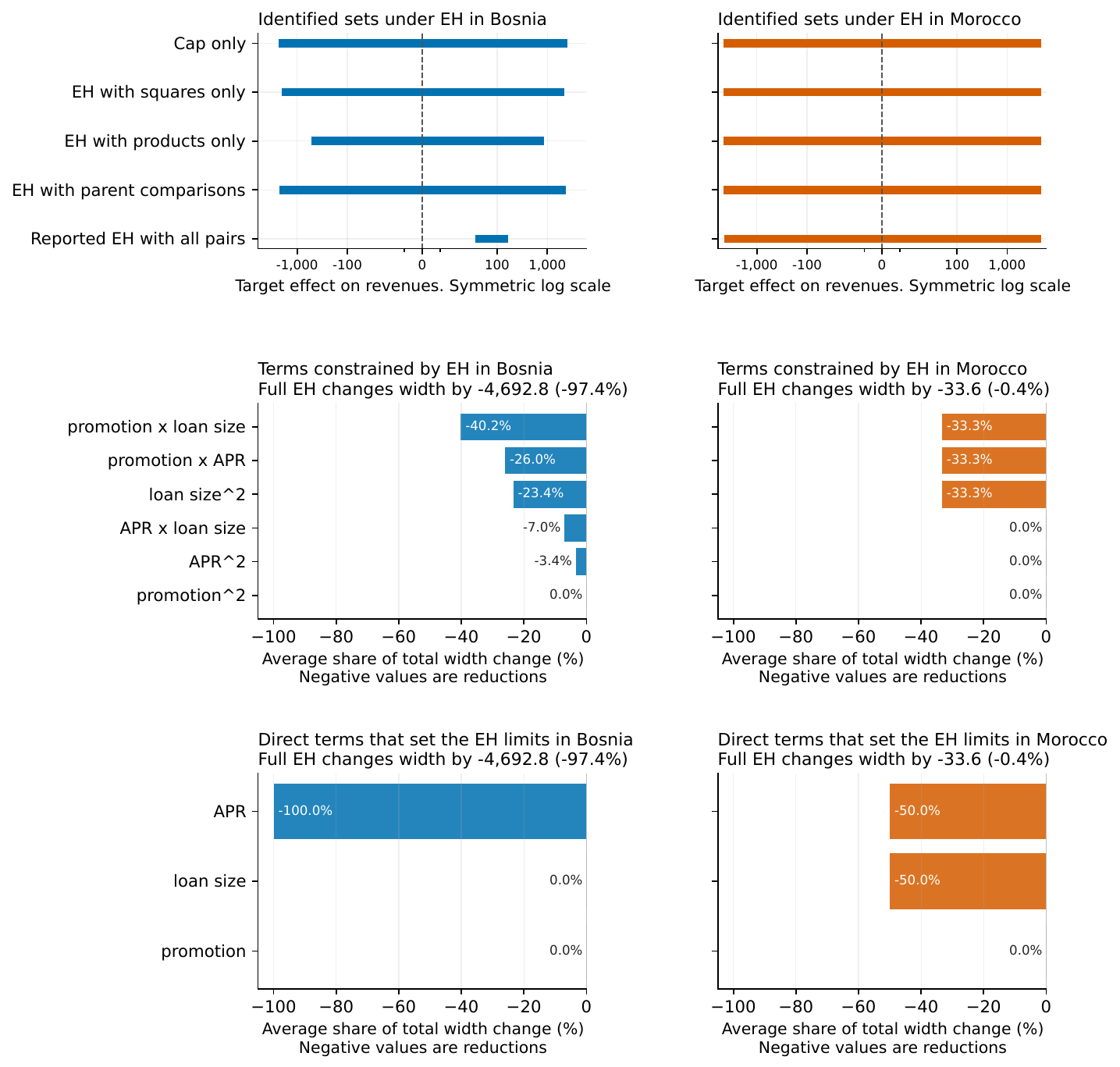}
  \caption{Where effect hierarchy binds.
  We use the magnitude cap in these panels and do not impose the APR sign
  restriction. In the top row, we compare four versions of effect hierarchy
  with the magnitude cap. In the middle row, we divide the width change under
  the reported rule among the six terms that effect hierarchy constrains. In
  the bottom row, we show the portion each direct term contributes to the limits.
  Each percentage is an average contribution across every order in which the
  comparisons can be added. Negative percentages mean reductions in width.}
  \label{fig:case-effect-ordering-diagnostics}
\end{figure}

In the top row of \cref{fig:case-effect-ordering-diagnostics}, we find that the
Bosnia result depends on the reported version of effect hierarchy. Under the reported rule, we compare every higher-order term with every direct
term in the same treatment arm. This rule reduces the width by 97.4\% and gives the positive
interval $[37.1,\,163.1]$ before we add the APR sign restriction. If effect hierarchy is applied in a parent-specific style, where the restriction is only applied to interactions against their ``parent'' direct terms (i.e., $|\beta_{\mathrm{APR \times Loan}}| \leq |\beta_{\mathrm{APR}}|$ and $|\beta_{\mathrm{APR \times Loan}}| \leq |\beta_{\mathrm{Loan}}|$ only), the width is reduced by only 4.3\% and the sign remains unresolved.

In the lower panels, we identify which comparisons produce this change. In Bosnia,
direct APR is the smallest direct coefficient in both arms at both endpoints.
It therefore sets the limit for every higher-order term. This use of APR is
separate from the APR sign restriction in \cref{fig:case-pair-bosnia-morocco}. Here we use the
magnitude of the APR coefficient as a limit, but we do not restrict its sign. Promotion by
loan size has the largest average share of the width change among the higher-order terms, at $-40.2\%$. The effect hierarchy principle compares this interaction with
direct APR even though APR is not part of the interaction. Under a parent-specific rule, we would not make this comparison, which explains why the parent-specific rule does not identify a sign. Overall, this indicates that if we believe that interaction terms are dominated by direct effects, then the Bosnia result is supported by the data. When the comparison is limited to parent direct effects, the sign remains unresolved.

Morocco is different. The reported rule reduces its width by only $33.6$
revenue units, or $0.4\%$. The control and treatment coefficients can move
together while their difference changes very little. We use this difference
to calculate the target treatment effect, so the interval also changes very
little. In the lower right panel, direct APR and loan size each have a share of
$-50\%$ of this small reduction. Promotion has a share of zero. In these bars,
we assign the reduction across the direct terms that set the effect hierarchy
limits. The bars do not
describe the APR sign restriction. The Bosnia result therefore depends on the strong comparison across all
pairs of terms.

\subsection{Practical guidance and robustness checks}
\label{subsec:robustness}

\begin{figure}[tbp]
  \centering
  \includegraphics[width=0.82\textwidth]{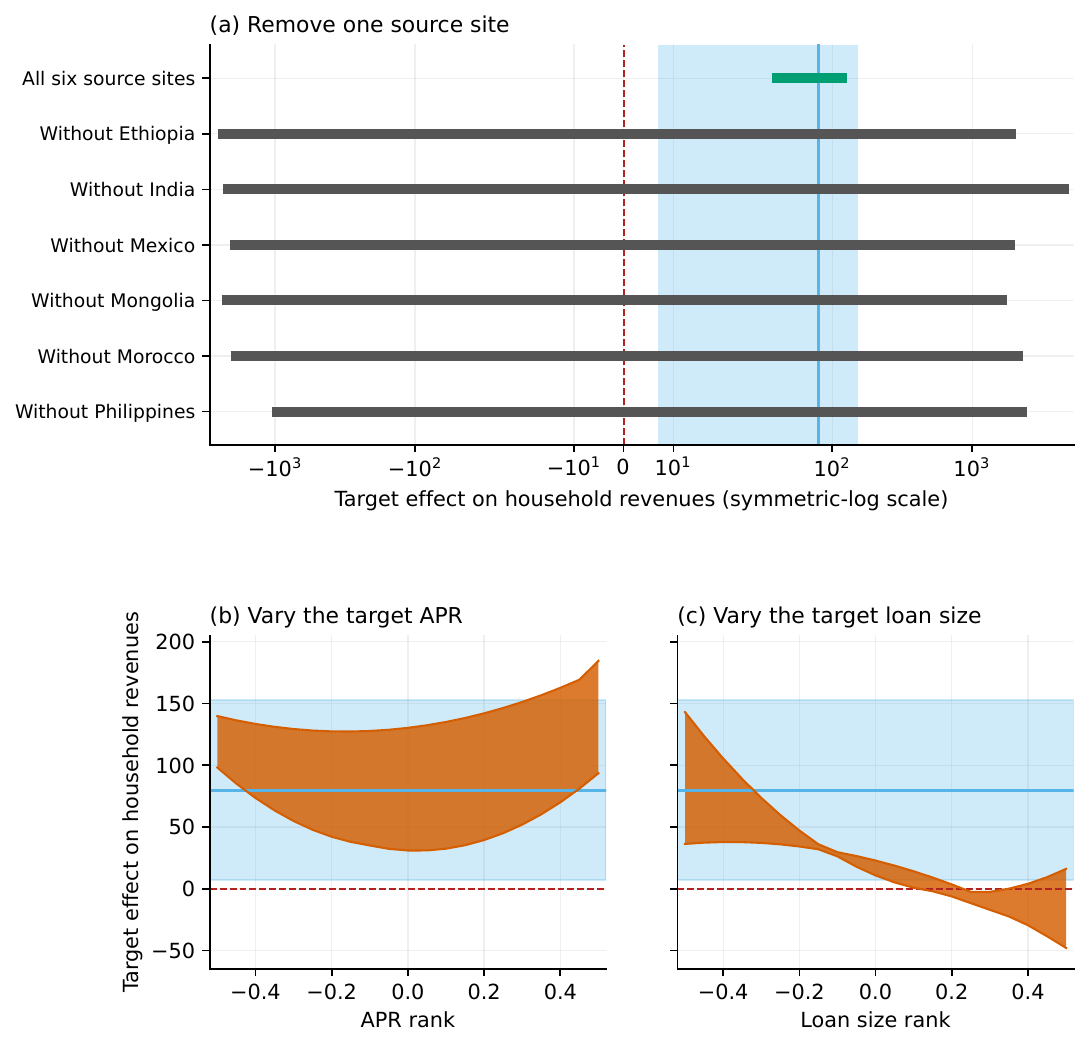}
  \caption{Sensitivity of Bosnia's identified set to source site removal and
    the target profile. Panel (a) removes one source site at a time after all
    restrictions. Panels (b) and (c) vary Bosnia's APR and loan size ranks
    across the range observed at the source sites. Thick bars and orange
    regions show outer identified set envelopes, and blue bands show the
    held-out experiment's 95\% interval. Sampling intervals are omitted so the
    figure isolates sensitivity of the identified set. The horizontal scale in
    panel (a) is symmetric log. Removing any source site makes the sign
    unresolved, whereas the sign survives most changes to APR and loan size.
    The conclusion therefore depends more on source site variation than on
    nearby target profiles.}
  \label{fig:case-bosnia-sensitivity}
\end{figure}

Sensitivity analysis should take care to separate conclusions driven by data from those that come solely from assumptions. We recommend a set of best practices that range from rigorous procedural guidelines to additional robustness checks that help evaluate the extent to which conclusions are sensitive to model assumptions and guard against overzealous use of the mathematical programming framework. 

We recommend assessing each restriction against a common baseline before combining restrictions. This distinguishes conclusions supported by a single assumption from those that require several assumptions jointly. Adding restrictions one at a time and reporting the bounds at each step produces a ladder that shows what each restriction contributes. \cref{fig:case-pair-bosnia-morocco} provides an example of such a ladder, using the magnitude cap as the baseline restriction.  Of course, the contribution of each restriction in this ladder may depend on which restrictions have already been imposed. To that end, we also recommend reconstructing the ladder with different orderings of the restrictions to see how the contribution of each restriction changes. Additionally, restrictions and plausible ranges for sensitivity parameters should be chosen before running the analysis. Ideally, these parameters should be chosen with a substantive justification in mind. For instance, the magnitude cap in \cref{fig:case-pair-bosnia-morocco} is generated from one standard deviation of revenues. Other recommendations include reporting results across the chosen sensitivity range, including where conclusions change, rather than selecting a single favorable specification.

More complex restrictions also warrant examining their constituent parts. The mathematical programming framework lends itself to conducting such comparisons through varying the set of constraints. \cref{fig:case-effect-ordering-diagnostics}  illustrates this approach for effect hierarchy; for aggregate budgets, one could relax the budget weights on selected groups of terms and recompute the bounds. Throughout these comparisons, the site variables, standardizations, and other parameters should remain fixed so that changes can be attributed to the restriction being examined. Infeasible and unbounded specifications should also be reported, as they reveal where restrictions conflict with the estimated evidence or leave target extrapolations uncontrolled.

Additional robustness checks can vary other aspects of the model or data to assess the stability of the conclusions. \cref{fig:case-bosnia-sensitivity} considers how changes in the underlying source data and target profile affect the identified set. The top panel asks whether one source
experiment is dispensable for the Bosnia conclusion. We remove each source
site in turn while holding the source rank transformation, target profile, and
restriction set at their values from the full set of six source sites. Every
removal lowers the span rank from five to four and produces an interval that
contains zero. No removal reverses the sign. Instead, the intervals become too
wide to determine it. Bosnia's positive identified set therefore uses
variation contributed by all six source sites. Reporting the results for each subset of source sites helps researchers understand how much the conclusion depends on the observed source variation and can be especially important when the number of source sites is small.

The lower panels ask whether the result is an accident of Bosnia's program
terms. We move its APR and loan size ranks across the full range observed at
the source sites, holding the source data and all other target features fixed.
The identified set stays positive at all 21 APR values and at 16 of 21 loan
size values. It loses the sign only toward the upper end of the loan size
range, far from Bosnia's observed program, whose loan size lies near the bottom
of the source site distribution. Thus, modest changes to Bosnia's target
profile do not explain the result. Rather, the diversity of source
sites does. This check can help assess how sensitive conclusions are to the target profile.

In our application, we find that in Bosnia effect hierarchy yields
positive estimated bounds that persist across plausible changes to the
target profile. However, the result depends on every source site and
is primarily driven by the effect hierarchy assumption applied to interaction terms. The program is infeasible in four of seven sample splits, and sampling uncertainty remains large. In Morocco, the same restrictions leave the sign unresolved. These checks clarify which assumptions and source evidence a conclusion depends on, and where the available information remains insufficient.

\FloatBarrier

\section{Discussion}
\label{sec:discussion}

External validity cannot be established in general from source experiments
alone. Learning about a new setting requires assumptions about how treatment
effects change across settings, leaving practitioners with an uncomfortable
choice: trust the source effects to carry over or give up on drawing a
conclusion about the target. We believe sensitivity analysis provides a
middle ground. Researchers can state which departures from the source
settings they consider plausible and examine which conclusions survive
those departures.

Our approach is one way to turn a modeler's substantive prior beliefs and
domain knowledge into quantifiable partial identification bounds. Parameter
restrictions describe which responses to changes in sites are plausible;
the source experiments then limit the target effects consistent with those
restrictions. Varying the restrictions makes their role in a conclusion
explicit. A conclusion that persists across plausible specifications has a
different evidential basis from one that requires a particular assumption.
The restrictions we study illustrate this use of sensitivity analysis, but
many other ways of representing knowledge about external validity remain
possible.

Developing useful parameter restrictions requires practitioners who
understand the mechanisms through which site-level variables change treatment effects.
Different applications may call for restrictions on the direction or
magnitude of those changes, or on how site-level features interact. Source
experiments and observational data could inform both the form and the
strength of these restrictions. Developing such restrictions, assessing
their substantive credibility, and accounting for uncertainty when they are
informed by data are promising directions for future research. This work
would make sensitivity analysis more useful by connecting what practitioners
know about an intervention to what they can conclude in a new setting.

\section*{Acknowledgements}
Egami acknowledges financial support from the National Science Foundation (SES-2318659).

\clearpage
\bibliographystyle{plainnat}
\bibliography{references}

\clearpage
\appendix
\crefalias{section}{appendix}
\crefalias{subsection}{appendix}
\crefalias{subsubsection}{appendix}

\section{Normal equations and identification of the target effect}
\label{app:identification}

We characterize the coefficients that reproduce the source conditional
means, then determine which target effects those coefficients permit.
This proves the normal equations \eqref{eq:profiled-moments} and
\cref{cor:point-id}. Under the sensitivity restrictions, the same
characterization gives the exact target-effect set. Proofs of the
supporting results appear in the subsections below.

\subsection{From the source equations to the target-effect set}
\label{subsec:normal-equations}

Profiling the unrestricted regression function leaves only variation in
context conditional on $X$. The following lemma characterizes what this
variation identifies. We assume throughout that the stated second moments
are finite. Write $\Ctil=C-\muC(X)$ and
$\dmom{a}=\CovR{a}=P\{\Ctil(\pseudo{a}-\mu_a(X))\}$ for the
source cross-moment in \eqref{eq:profiled-moments}.

\begin{lemma}[Profiled loss and source-compatible coefficients]
\label{lem:profiled-source-characterization}
For each arm $a$ and coefficient $\beta$, the profiled minimizer is
$f_{a,\beta}(x)=\mu_a(x)-\muC(x)^\top\beta$. The coefficients minimizing
the profiled loss are exactly the solutions of
$\CovC\beta=\dmom{a}$, and $\dmom{a}\in\range(\CovC)$.
Under the response model \eqref{eq:app-response-model}, these are exactly
the coefficients for which
\[
  f_{a,\beta}(X)+C^\top\beta=\mu_a(X,C)
  \quad P\text{-almost surely}.
\]
\textnormal{\hyperref[subsec:identification-proofs]{Proof in Appendix~\ref*{subsec:identification-proofs}.}}
\end{lemma}

\Cref{lem:profiled-source-characterization} gives
\eqref{eq:profiled-moments}. To determine the target effect, impose target
response invariance,
\begin{equation}
  \E[Y_T(a)\mid X_T=x]=g_a(x)+\ct^\top\beta_a,
  \qquad a\in\{0,1\},
  \label{eq:app-target-response-model}
\end{equation}
and $Q_T\ll P_X$. Substituting the profiled functions into this target
mean gives
\[
  \psi_T=Q_T(\mu_1-\mu_0)
    +\{\ct-Q_T\muC\}^\top(\beta_1-\beta_0)
  =b_T+\ellvec^\top(\beta_1-\beta_0).
\]
Thus the remaining identification question is whether the source equations
fix this linear functional of the coefficients.

\phantomsection\label{proof:point-id}
To prove \cref{cor:point-id}, use $\dmom{a}\in\range(\CovC)$, so $\CovC\CovC^+\dmom{a}=\dmom{a}$. Thus, $\CovC^+\dmom{a}$ is one solution, and subtracting it from any other solution gives an element of $\kernel(\CovC)$. Consequently,
$\{\beta_a:\CovC\beta_a=\dmom{a}\}
=\CovC^+\dmom{a}+\kernel(\CovC)$. Substitution into the target effect gives
\[
  \psi_T=b_T+\ellvec^\top\CovC^+(\dmom{1}-\dmom{0})
  +\ellvec^\top(z_1-z_0).
\]
The final term vanishes for every $z_0,z_1\in\kernel(\CovC)$ exactly when
$\ellvec\in\kernel(\CovC)^\perp=\range(\CovC)$. If this condition fails,
choose $z\in\kernel(\CovC)$ with $\ellvec^\top z\ne0$ and replace $\beta_1$
by $\beta_1+tz$. Each pair therefore reproduces the same source
conditional means by \cref{lem:profiled-source-characterization}. The target effect ranges
over $\R$ as $t$ varies.

The range condition has a direct interpretation in terms of the observed
site profiles. Recall $\Vspan=\operatorname{span}\{c_s-c_1:s\geq2\}$
and its orthonormal basis $B_C$.

Both $C-c_1$ and its conditional expectation given $X$ lie in $\Vspan$,
so $\Ctil\in\Vspan$ almost surely. Writing $Z=B_C^\top\Ctil$ gives
$\Ctil=B_CZ$ and $\CovC=B_CP(ZZ^\top)B_C^\top$. Consequently,
\begin{equation}
  \range(\CovC)\subseteq\Vspan,
  \qquad \rankop(\CovC)\leq r\leq S-1,
  \qquad \dmom{a}\in\Vspan.
  \label{eq:id-rank-bounds}
\end{equation}
Under \eqref{eq:setup-within-span}, $P(ZZ^\top)=\Omega\succ0$, so
\begin{equation}
  \CovC=\Bspan\Om\Bspan^\top,
  \qquad \range(\CovC)=\Vspan,
  \qquad \kernel(\CovC)=\Vspan^\perp.
  \label{eq:id-q-factorization}
\end{equation}

To translate the range condition into a condition on the target profile,
let $p_s(x)=P(C=c_s\mid X=x)$. Covariate overlap gives
\[
  \bar c:=Q_T\muC=\sum_{s=1}^S c_s Q_Tp_s
  \in\convop\{c_1,\ldots,c_S\}\subseteq c_1+\Vspan.
\]
Since $\ellvec=\ct-\bar c$ and $\range(\CovC)=\Vspan$, we obtain
\begin{equation}
  \ellvec\in\range(\CovC)
  \quad\Longleftrightarrow\quad
  \ct\in\affop\{c_1,\ldots,c_S\}.
  \label{eq:id-target-affine-span}
\end{equation}
By \cref{cor:point-id}, the unrestricted target effect is therefore
$\bT+\ellvec^\top\CovC^+(\dmom{1}-\dmom{0})$ when the target profile
lies in this affine span, and is unbounded in both directions otherwise.
In particular, it is point identified whenever $\ct$ is a realized source
profile.

When this condition fails, sensitivity restrictions limit the coefficient
directions that the source cannot identify. Intersecting the normal
equations with $\restrset$ gives $\feasset$. The following proposition
shows that projecting this feasible set produces exactly the target
effects compatible with the model.

\begin{proposition}[Exact projection and mean sharpness]
\label{prop:mean-sharpness}
Assume the response model \eqref{eq:app-response-model}, target response invariance
\eqref{eq:app-target-response-model}, unrestricted support for the potential outcomes, and that
$\feasset$ is nonempty. The exact target-effect set is
\begin{equation}
  \mathcal I_T
  =\{\bT+\ellvec^\top(\beta_1-\beta_0):
  \beta\in\feasset\}.
  \label{eq:id-exact-projection}
\end{equation}
Every value in this set is attained by a source-target response model that leaves the observed source
law unchanged. Each polyhedral branch of $\feasset$ projects to an interval, possibly empty,
unbounded, or a singleton. A finite union projects to a finite union of such intervals, and
$[\vminus,\vplus]$ is its convex hull. For a convex restriction, the exact set equals this interval.
\textnormal{\hyperref[subsec:sharpness-proof]{Proof in Appendix~\ref*{subsec:sharpness-proof}.}}
\end{proposition}

Consequently, the optimization problems in \eqref{eq:endpoint-programs}
give the sharp lower and upper bounds. For a finite union, the branchwise
intervals retain any gaps in the exact set; the two outer bounds describe
its convex hull.

\subsection{Proof of \texorpdfstring{\hyperref[lem:profiled-source-characterization]{Lemma~\ref*{lem:profiled-source-characterization}}}{Lemma \ref*{lem:profiled-source-characterization}}}
\label{subsec:identification-proofs}
Fix an arm $a$ and a coefficient vector $\beta$. Conditional expectation is
the $L_2$ projection on functions of $X$, so profiling $f$ from
$E\{R_a-f(X)-C^\top\beta\}^2$ gives
\[
  f(x)=\mu_a(x)-E(C\mid X=x)^\top\beta.
\]
The residual context is
\begin{equation}
  \muC(x)=\E_P[C\mid X=x],
  \qquad
  \Ctil=C-\muC(X),
  \label{eq:app-residual-context}
\end{equation}
and the source moments are
\begin{equation}
  \CovC=P(\Ctil\Ctil^\top),
  \qquad
  \dmom{a}=P\{\Ctil(\pseudo{a}-\mu_a(X))\}.
  \label{eq:app-source-moments}
\end{equation}
Throughout the appendices, $\CovC$ and $\dmom{a}$ abbreviate the covariances written in operator form
in the main text: because $\Ctil$ is centered, $\CovC=\E_P[\cov(C\mid X)]$ and $\dmom{a}=\CovR{a}$.
The moment $\dmom{a}$ is unchanged if $\pseudo{a}$ is replaced by $R_a$. After
profiling, the part of the loss that depends on $\beta$ is
\[
  -2\beta^\top\dmom{a}+\beta^\top\CovC\beta.
\]
Its normal equation is therefore $\CovC\beta=\dmom{a}$.

For completeness, $\CovC$ is positive semidefinite because
\begin{equation}
  z^\top\CovC z=E\{(z^\top\Ctil)^2\}\geq0.
  \label{eq:id-q-positive-semidefinite}
\end{equation}
If $z\in\kernel(\CovC)$, then $z^\top\Ctil=0$ almost surely and hence
$z^\top\dmom{a}=0$. Thus $\dmom{a}\in\range(\CovC)$, using
\[
  \range(\CovC)=\kernel(\CovC)^\perp
\]
for a symmetric matrix. Completing the quadratic around any solution shows
that the minimizers are exactly the solutions of the normal equation.

Randomization and consistency imply
\begin{equation}
  \E_P[R_a\mid X,C]=\mu_a(X,C)
  =g_a^0(X)+C^\top\beta_a^0
  \label{eq:id-true-conditional-mean}
\end{equation}
for some $(g_a^0,\beta_a^0)$ satisfying \eqref{eq:app-response-model}. Taking the conditional
expectation given $X$ gives
\begin{equation}
  \mu_a(X)=g_a^0(X)+\muC(X)^\top\beta_a^0.
\end{equation}
Subtracting the last display from \eqref{eq:id-true-conditional-mean},
multiplying by $\Ctil$, and taking expectations gives
\begin{equation}
\begin{split}
  \dmom{a}
  &=P\{\Ctil(\mu_a(X,C)-\mu_a(X))\}\\
  &=P(\Ctil\Ctil^\top)\beta_a^0
  =\CovC\beta_a^0.
  \label{eq:id-true-coefficient-solves-foc}
\end{split}
\end{equation}
If
$\CovC\beta=\dmom{a}$, then $z=\beta-\beta_a^0$ satisfies $\CovC z=0$, and
\eqref{eq:id-q-positive-semidefinite} gives
$z^\top\Ctil=0$ almost surely. With $g_a=f_{a,\beta}$,
\begin{equation}
\begin{split}
  g_a(X)+C^\top\beta
  =\mu_a(X)+\Ctil^\top\beta
  =\mu_a(X)+\Ctil^\top\beta_a^0
  =\mu_a(X,C)
\end{split}
\end{equation}
on the source support. Thus the solutions of \eqref{eq:profiled-moments}
are exactly the coefficients that reproduce the source conditional means
when paired with their profiled regression functions.
The argument uses only
$E(R_a\mid X,C)=E\{Y(a)\mid X,C\}$, so it also applies when the known,
positive assignment probability varies with $(X,C)$.

\subsection{Proof of \texorpdfstring{\hyperref[prop:mean-sharpness]{Proposition~\ref*{prop:mean-sharpness}}}{Proposition \ref*{prop:mean-sharpness}}}
\label{subsec:sharpness-proof}
We prove necessity and attainability separately.

\paragraph{Necessity.}
Consider any source-target response model admitted by the assumptions
and the restriction $\restrset$. Its coefficients satisfy
$\CovC\beta_a=\dmom{a}$ by
\cref{lem:profiled-source-characterization}. They also satisfy
$\beta\in\restrset$ by construction. Hence
$\beta\in\feasset$. Target response invariance
\eqref{eq:app-target-response-model} and
\eqref{eq:app-target-effect} then imply
\begin{equation}
  \E\{Y_T(1)-Y_T(0)\}
  =\bT+\ellvec^\top(\beta_1-\beta_0).
\end{equation}
Every admissible target effect therefore belongs to the right side of
\eqref{eq:id-exact-projection}.

\paragraph{Attainability.}
Fix an arbitrary $\beta\in\feasset$ and define
\begin{equation}
  g_a(x)=\mu_a(x)-\muC(x)^\top\beta_a.
  \label{eq:id-sharp-g}
\end{equation}
Because $\CovC\beta_a=\dmom{a}$ for both arms,
\cref{lem:profiled-source-characterization} gives
\begin{equation}
  g_a(X)+C^\top\beta_a=\mu_a(X,C)
  \quad P\text{-almost surely},
  \label{eq:id-sharp-source-mean}
\end{equation}
on the source support.

We now construct a full source potential-outcome law. Let
\begin{equation}
  F_a(\mathcal B\mid x,c)
  =P(Y\in\mathcal B\mid A=a,X=x,C=c)
\end{equation}
denote the observed conditional outcome law in arm $a$. Draw $(X,C)$
from its observed source marginal. Conditional on $(X,C)=(x,c)$, draw
$(Y(0),Y(1))$ from any coupling whose two marginal laws are
$F_0(\cdot\mid x,c)$ and $F_1(\cdot\mid x,c)$. The conditional product
of these laws is one such coupling. Next draw $A$ independently of
$(Y(0),Y(1))$ given $(X,C)$, using the known probabilities
$\propen{a}(X,C)$, and set
\begin{equation}
  Y=AY(1)+(1-A)Y(0).
\end{equation}
This construction satisfies consistency and conditional randomization.
It also preserves the complete observed source law. To see this, for
every measurable $\mathcal B$ and each arm $a$,
\begin{equation}
\begin{split}
  &P^*(Y\in\mathcal B,A=a\mid X=x,C=c)\\
  &\qquad
  =P^*(A=a\mid X=x,C=c)
    P^*(Y(a)\in\mathcal B\mid X=x,C=c)\\
  &\qquad
  =\propen{a}(x,c)F_a(\mathcal B\mid x,c),
  \label{eq:id-sharp-observed-law}
\end{split}
\end{equation}
which is the corresponding conditional probability under the observed
law $P$. The marginal law of $(X,C)$ was also kept fixed, so
$P^*=P$ on $(Y,A,X,C)$.

The mean of $F_a(\cdot\mid x,c)$ is $\mu_a(x,c)$. Therefore the constructed
potential outcomes satisfy
\begin{equation}
\begin{split}
  \E_{P^*}[Y(a)\mid X,C]
  &=\mu_a(X,C)\\
  &=g_a(X)+C^\top\beta_a,
\end{split}
\end{equation}
where the second equality is \eqref{eq:id-sharp-source-mean}. Thus the
constructed source law obeys the response model with the chosen
coefficient pair even though the observed law is unchanged.

It remains to construct the target law. Draw $X_T\sim\tP$; the function
$g_a$ in \eqref{eq:id-sharp-g} is defined $\tP$-almost everywhere
because $\tP\ll P_X$. Define
\begin{equation}
  Y_T(a)=g_a(X_T)+\ct^\top\beta_a+\varepsilon_{a,T},
\end{equation}
where $(\varepsilon_{0,T},\varepsilon_{1,T})$ has conditional mean zero
given $X_T$ and finite second moments. One may take both residuals to be
zero. The unrestricted outcome-support assumption permits this target
construction. Conditional expectation gives, arm by arm,
\begin{equation}
  \E[Y_T(a)\mid X_T]
  =g_a(X_T)+\ct^\top\beta_a,
\end{equation}
so target response invariance \eqref{eq:app-target-response-model} holds for the constructed law. Its
average treatment effect is
\begin{equation}
\begin{split}
  \E\{Y_T(1)-Y_T(0)\}
  &=\tP(g_1-g_0)+\ct^\top(\beta_1-\beta_0)\\
  &=\tP\{\mu_1-\mu_0-\muC^\top(\beta_1-\beta_0)\}
    +\ct^\top(\beta_1-\beta_0)\\
  &=\tP(\mu_1-\mu_0)
    +\{\ct-\tP\muC\}^\top(\beta_1-\beta_0)\\
  &=\bT+\ellvec^\top(\beta_1-\beta_0).
  \label{eq:id-sharp-target-effect}
\end{split}
\end{equation}
The second equality substitutes \eqref{eq:id-sharp-g}; the last equality
uses \eqref{eq:app-target-coefficients}. Since the initial
$\beta\in\feasset$ was arbitrary, every value on the right
side of \eqref{eq:id-exact-projection} is attained. Combining this
construction with the necessity argument proves equality in
\eqref{eq:id-exact-projection}.

Finally, Fourier--Motzkin elimination implies that
\[
  \{\Psi_0+q^\top x:x\in\mathcal P\}
  \text{ is a closed interval, half-line, singleton, }\R,\text{ or empty}
\]
for any polyhedron $\mathcal P$. Taking the finite union proves the branch
claim. If $\mathcal F$ is convex, its affine image is convex and therefore
equals the interval between its two extrema.

\section{Proof of \texorpdfstring{\hyperref[thm:endpoint-clt]{Theorem~\ref*{thm:endpoint-clt}}}{Theorem 1}}
\label{app:setup}
\label{subsec:proof-theorem-three}
\label{app:influence}

Fix an endpoint $j\in\{-,+\}$. We first consider its population-winning
polyhedron. Use the moment vector
$\Gamma=(b_T,\ellvec,\CovC,\dmom{0},\dmom{1})$ and the reduced coordinates
$\gamma=(b,\ell,\Omega,\varsigma_0,\varsigma_1)$ of
\eqref{eq:site-moment-domain}. The basis $B_C$ is fixed by the source
profiles. Write $\Ctil=C-\muC(X)$ and
$\widetilde U_a=\pseudo{a}-\mu_a(X)$, where $\pseudo{a}$ is the augmented outcome in
\eqref{eq:augmented-outcome}.

The first step is to identify the sampling error that will perturb the
program. Define the source gradients by
\begin{align*}
  \varphi_\Sigma&=\Ctil\Ctil^\top-\CovC,&
  \varphi_{d_a}&=\Ctil \widetilde U_a-\dmom{a},\\
  \varphi_b&=w(X)(\widetilde U_1-\widetilde U_0),&
  \varphi_\ell&=-w(X)\Ctil,
\end{align*}
and the target gradients by
\[
  \varphi_{T,b}=\mu_1(X_T)-\mu_0(X_T)-b_T,\qquad
  \varphi_{T,\ell}=Q_T\muC-\muC(X_T),
\]
with zero target components for $(\CovC,\dmom{0},\dmom{1})$. Stack these
in the order of $\Gamma$ as $\varphi_\Gamma$ and
$\varphi_{T,\Gamma}$. Conditional centering of $\Ctil$ and $\widetilde U_a$ makes
all components mean zero.

\begin{lemma}[Cross-fitted input expansion]
\label{prop:primitive-gradients}
Under \cref{ass:sampling,ass:nuisance-plugin}, \cref{alg:crossfit-plugin} satisfies
\begin{equation}
  \what\Gamma-\Gamma
  =\frac1n\sum_{i=1}^n\varphi_\Gamma(O_i)
  +\frac1{n_T}\sum_{i=1}^{n_T}\varphi_{T,\Gamma}(X_{T,i})
  +o_p(n^{-1/2}).
  \label{eq:key-input-expansion}
\end{equation}
The same expansion holds after applying the fixed linear reduction
\[
  (b,\ell,\Sigma,d_0,d_1)\mapsto
  (b,\ell,\Bspan^\top\Sigma\Bspan,
  \Bspan^\top d_0,\Bspan^\top d_1).
\]
The empirical moments belong to $\mathcal G_C$ with probability tending to
one. If $Q_T$ is known and integrated exactly, the target sum is absent.
\textnormal{\hyperref[subsec:input-expansion-proof]{Proof in Appendix~\ref*{subsec:input-expansion-proof}.}}
\end{lemma}

Consequently, $\what\gamma-\gamma^*=O_p(n^{-1/2})$, and the estimator
perturbs the program within the fixed-span domain on which we can
differentiate its value. This restriction is needed because an arbitrary
perturbation of the singular full equality system could make it infeasible.
The next result gives the change in the optimized value along the
perturbations that the estimator actually produces.

\begin{lemma}[Derivative of the partial identification bound]
\label{prop:value-derivative}
Suppose \cref{ass:regular-endpoint} holds for the branch. Its endpoint value
$v_j$ on $\mathcal G_C$ is Hadamard differentiable at $\Gamma$ tangentially
to the tangent space of $\mathcal G_C$. For a tangent perturbation
$\dot\Gamma=(\dot b,\dot\ell,\dot\Sigma,\dot d_0,\dot d_1)$,
\[
  Dv_j(\Gamma)[\dot\Gamma]
  =\dot b+\dot\ell^\top(\beta^*_{j,1}-\beta^*_{j,0})
  +\sum_{a\in\{0,1\}}\lambda_{j,a}^\top
  (\dot d_a-\dot\Sigma\beta^*_{j,a}).
\]
Here $\lambda_{j,a}=B_C\bar\lambda_{j,a}$ is the minimum-norm full
equality dual, with the upper-endpoint dual equal to the negative of the dual for
the reflected minimization. Every full dual representative gives the
same derivative. For a finite union, \cref{ass:branch-separation} makes the union
value agree locally with its population-winning branch, so it has the same
derivative.
\textnormal{\hyperref[subsec:proof-proposition-two]{Proof in Appendix~\ref*{subsec:proof-proposition-two}.}}
\end{lemma}

With probability tending to one, $\what\gamma$ lies in the regular
neighborhood of the population input. The finite-dimensional delta method
\citep[Theorem~3.9.4]{van1996weak}, applied in the open reduced coordinates,
therefore gives
\begin{equation}
  \what v_j-v_j
  =Dv_j(\Gamma)[\what\Gamma-\Gamma]+o_p(n^{-1/2}).
  \label{eq:central-value-expansion}
\end{equation}
For a finite union the local equality in \cref{prop:value-derivative}
puts the union estimator under the same expansion.

Applying the derivative to the source gradient gives \eqref{eq:source-if}.
In particular, the constraint contribution simplifies to
\[
  \varphi_{d_a}-\varphi_\Sigma\beta^*_{j,a}
  =\Ctil\{\widetilde U_a-\Ctil^\top\beta^*_{j,a}\},
\]
because $\CovC\beta^*_{j,a}=\dmom{a}$. The derivative applied to the
target gradient gives \eqref{eq:target-if}. Substituting
\eqref{eq:key-input-expansion} into \eqref{eq:central-value-expansion}
thus yields
\[
  \sqrt n(\what v_j-v_j)
  =\frac1{\sqrt n}\sum_{i=1}^n\phi_j(O_i)
  +\frac{\sqrt n}{n_T}\sum_{i=1}^{n_T}\phi_{T,j}(X_{T,i})+o_p(1).
\]
The two sums are independent and centered. Their finite $2+\epsilon$
moments and $n_T/n\to\zeta\in(0,\infty]$ imply the Lindeberg condition,
so their sum converges to a centered normal law with variance
\[
  V_j=\var_P\{\phi_j(O)\}
       +\zeta^{-1}\var_{Q_T}\{\phi_{T,j}(X_T)\}.
\]
When $\zeta=\infty$, the target sum converges to zero in probability.
This proves the distributional conclusion of the theorem.

It remains to justify replacing the population influence functions by the
reported estimates. In addition to fitted nuisances, these estimates use
an optimizer and dual variables computed from the whole moment vector,
so their consistency must also be established.

\begin{lemma}[Consistency of the reported influence-function variance]
\label{lem:reported-variance}
Under the assumptions of \cref{thm:endpoint-clt}, the reported optimizer
and reduced dual converge to their population values. The minimum-norm
full dual converges as well. Replacing these quantities and the nuisance
functions in the source and target influence functions changes their
empirical first and second moments by $o_p(1)$. Consequently,
$\what V_j\cp V_j$.
\textnormal{\hyperref[subsec:reported-variance-proof]{Proof in Appendix~\ref*{subsec:reported-variance-proof}.}}
\end{lemma}

\Cref{lem:reported-variance} proves the variance conclusion. If $V_j>0$,
Slutsky's theorem also justifies the reported Wald intervals. If $Q_T$ is
known, exact integration removes the target empirical process in
\cref{prop:primitive-gradients}; the same derivative and variance argument
then give every conclusion with the target sum and variance term deleted.

\subsection{Proof of \texorpdfstring{\hyperref[prop:primitive-gradients]{Lemma~\ref*{prop:primitive-gradients}}}{Cross-fitted input expansion}}
\label{subsec:input-expansion-proof}
Condition on the training samples for one fold and suppress its index.
Write $P_k$ and $Q_k$ for the held-out source and target empirical
measures, and define the nuisance errors
\[
  e_C=\what m_C-\muC,\qquad e_{\mu,a}=\what\mu_a-\mu_a,\qquad
  e_w=\what w-w,\qquad
  e_{U,a}=\what{\pseudo{a}}-\pseudo{a}.
\]
Here $e_{\mu,a}$ is the error of the marginal regression on $X$. If
$e_{q,a}(X,C)=\what\mu_a(X,C)-\mu_a(X,C)$ is the joint regression error,
known randomization gives
\[
  e_{U,a}
  =\left\{1-\frac{\one\{A=a\}}{\propen{a}(X,C)}\right\}e_{q,a}(X,C),
  \qquad E(e_{U,a}\mid X,C)=0.
\]
Thus $e_{U,a}$ is $L_2(P)$ consistent and uniformly $L_q(P)$ bounded.
Also $E(\Ctil\mid X)=E(\widetilde U_a\mid X)=0$.

Substitute $\what\Ctil=\Ctil-e_C$ and
$\what{\widetilde U}_a=\widetilde U_a+e_{U,a}-e_{\mu,a}$ into the scores of
\eqref{eq:orthogonal-source-moments}--\eqref{eq:orthogonal-target-moments}.
For each component $Z$ of $\Gamma$, this gives the exact fold identity
\begin{equation}
  \what Z^{(k)}-Z
  =(P_k-P)\varphi_Z+(Q_k-Q_T)\varphi_{T,Z}
    +D_Z+(P_k-P)H_Z+(Q_k-Q_T)J_Z,
  \label{eq:aux-fold-expansion}
\end{equation}
where terms with zero target gradient are omitted. The population drifts
are
\begin{align*}
  D_\Sigma&=P(e_Ce_C^\top),&
  D_{d_a}&=P(e_Ce_{\mu,a}),\\
  D_b&=-P\{e_w(e_{\mu,1}-e_{\mu,0})\},&
  D_\ell&=P(e_we_C).
\end{align*}
Conditional centering removes the linear errors in the source covariance
and cross-moments. For the transported terms, $Q_Te_{\mu,a}=P(we_{\mu,a})$ and
$Q_Te_C=P(we_C)$ cancel the linear target errors against the source
corrections. To make the empirical remainders explicit,
put
\[
  H_{\theta_a}=e_w\widetilde U_a+we_{U,a}-we_{\mu,a}+e_we_{U,a}-e_we_{\mu,a}.
\]
Then
\begin{align*}
  H_\Sigma&=-\Ctil e_C^\top-e_C\Ctil^\top+e_Ce_C^\top,\\
  H_{d_a}&=\Ctil e_{U,a}-\Ctil e_{\mu,a}-e_C\widetilde U_a-e_Ce_{U,a}+e_Ce_{\mu,a},\\
  H_b&=H_{\theta_1}-H_{\theta_0},&J_b&=e_{\mu,1}-e_{\mu,0},\\
  H_\ell&=-e_w\Ctil+we_C+e_we_C,&J_\ell&=-e_C,
\end{align*}
with $J_\Sigma=J_{d_a}=0$. These formulas also verify the signs of the
four drifts.

Cauchy--Schwarz and \eqref{eq:product-rates} give
$\|D_Z\|=o_p(n^{-1/2})$ for every component. For the empirical
remainders, interpolation between $L_2$ and $L_q$, with $q>4$, makes
every nuisance error $o_p(1)$ in $L_4(P)$. Each term in $H_Z$ is a
product with at least one such error and with its other factor bounded in
$L_4(P)$. H\"older's inequality therefore gives
$\|H_Z\|_{P,2}=o_p(1)$. The target consistency assumptions similarly
give $\|J_Z\|_{Q_T,2}=o_p(1)$.

For a function $H$ fixed conditional on training, independence of the
holdout gives
\[
  E\left[\left\|\sqrt{n_k}(P_k-P)H\right\|^2
       \,\middle|\,\text{training}\right]\leq P\|H\|^2.
\]
Conditional Chebyshev, first on events where the right side is small and
then unconditionally, shows that the source remainders are
$o_p(n_k^{-1/2})$. The same argument gives
$o_p(n_{T,k}^{-1/2})$ for the target remainders. With a fixed number of
balanced folds and $n_T/n\to\zeta>0$, both are $o_p(n^{-1/2})$.

Pool the source and target terms with their respective observation
weights. The leading empirical terms become the two full-sample sums in
\eqref{eq:key-input-expansion}. The source and target fold weights can
differ by $O(n^{-1}+n_T^{-1})=O(n^{-1})$; their effect on the canceled
population terms is $o_p(n^{-1/2})$, since there are finitely many folds
and those terms are bounded in probability. The drifts and empirical
remainders remain $o_p(n^{-1/2})$ after pooling. This proves the joint
expansion, and its reduced version follows by the fixed linear map.

Finally, \cref{alg:crossfit-plugin} estimates $m_C$ through joint site probabilities, so
$C-\what m_C(X)\in\Vspan$ for every held-out observation. Hence
\[
  \what\CovC=B_C\what\Omega B_C^\top,\qquad
  \what d_a=B_C\what\varsigma_a.
\]
The expansion implies $\what\Omega\to_p\Omega$, and the smallest
population eigenvalue is bounded away from zero. Weyl's inequality gives
$\what\Omega\succ0$ with probability tending to one, which places
$\what\Gamma$ in $\mathcal G_C$. If $Q_T$ is known, replace each $Q_k$
by $Q_T$ above. All target empirical terms disappear, while the same
population cancellations and remainder bounds continue to hold.

\subsection{Proof of \texorpdfstring{\hyperref[prop:value-derivative]{Lemma~\ref*{prop:value-derivative}}}{Derivative of the bound}}
\label{subsec:proof-proposition-two}
\label{app:optimization}

Use the fixed source span $\Vspan$ in \eqref{eq:source-context-span} and
its orthonormal basis $B_C$. Under the within-span condition
\begin{equation}
  \Omega=B_C^\top\CovC B_C\succ0,\qquad
  \lambda_{\min}(\Omega)\ge\kappaOm>0,
  \label{eq:setup-within-span}
\end{equation}
\eqref{eq:id-q-factorization} gives $\CovC=B_C\Omega B_C^\top$ and
$d_a=B_C\varsigma_a$, where $\varsigma_a=B_C^\top d_a$. We therefore work with
\begin{equation}
  \gamma=(b,\ell,\Omega,\varsigma_0,\varsigma_1)
  \label{eq:setup-reduced-input}
\end{equation}
in the open domain
\begin{equation}
  \mathcal U=\R\times\R^p\times\mathbb S_{++}^r\times\R^r\times\R^r.
  \label{eq:setup-open-domain}
\end{equation}
The fixed linear chart
\begin{equation}
  T(\gamma)=(b,\ell,B_C\Omega B_C^\top,B_C\varsigma_0,B_C\varsigma_1),
  \qquad \mathcal G_C=T(\mathcal U),
  \label{eq:setup-site-input-domain}
\end{equation}
identifies this domain with the admissible original moments in
\eqref{eq:site-moment-domain}.

For a restriction set
\begin{equation}
  \restrset=\bigcup_{r_0\in\mathcal J}\{\beta:G_{r_0}\beta\le h_{r_0}\},
  \label{eq:setup-restriction-union}
\end{equation}
fix a branch $G\beta\le h$ and stack
$\beta=(\beta_0^\top,\beta_1^\top)^\top$. Define
\begin{equation}
  c(\ell)=\begin{bmatrix}-\ell\\\ell\end{bmatrix},\qquad
  H(\Omega)=\begin{bmatrix}\Omega B_C^\top&0\\0&\Omega B_C^\top\end{bmatrix},
  \qquad e(\varsigma_0,\varsigma_1)=\begin{bmatrix}\varsigma_0\\\varsigma_1\end{bmatrix}.
  \label{eq:setup-branch-objects}
\end{equation}
The reduced lower-endpoint problem is
\begin{equation}
  \branchval(\gamma)=\inf_{\beta}
  \{b+c(\ell)^\top\beta:H(\Omega)\beta=e(\varsigma_0,\varsigma_1),\ G\beta\le h\}.
  \label{eq:setup-branch-program}
\end{equation}
\Cref{ass:regular-endpoint} makes its value finite and attained on a neighborhood
$\mathcal U_0$ of $\gamma^*$. Its Lagrangian is
\begin{equation}
  \mathcal L(\beta,\bar\lambda,\nu;\gamma)
  =b+c(\ell)^\top\beta+\bar\lambda^\top\{e-H(\Omega)\beta\}
    +\nu^\top(G\beta-h),
  \label{eq:setup-reduced-lagrangian}
\end{equation}
with $\bar\lambda\in\R^{2r}$ and $\nu\ge0$. Write
$\kkt=(\beta,\bar\lambda,\nu)$ and
$\kkt^*=(\beta^*,\bar\lambda^*,\nu^*)$ for a KKT triple and the unique
population triple, respectively. The value is $+\infty$ on infeasible
branches and may be $-\infty$ on unbounded branches outside the regular
neighborhood.

The reduction preserves the feasible coefficients: multiplying by $B_C^\top$
in one direction and by $B_C$ in the other gives
\begin{equation}
  \CovC\beta_a=\dmom{a}
  \quad\Longleftrightarrow\quad
  \Om\Bspan^\top\beta_a=\xired{a}.
  \label{eq:opt-row-equivalence}
\end{equation}
Moreover, $\Hred(\Om)\Hred(\Om)^\top=I_2\otimes\Om^2$, so
\begin{equation}
  \sigma_{\min}\{\Hred(\Om)\}=\lambda_{\min}(\Om)>0.
  \label{eq:opt-sigma-min}
\end{equation}
This singular value is at least $\kappaOm/2$ on a sufficiently small
neighborhood of the population input. Thus the reduced equalities have
full row rank even when the original covariance is singular.

\paragraph{Derivative in reduced coordinates.}
We now differentiate the reduced value. Write
$H_0=\Hred(\Om^\ast)$, $e_0=\ered(\varsigma_0^\ast,\varsigma_1^\ast)$,
$g=c(\ell^\ast)$, and use $H_n,e_n,g_n$ for the corresponding quantities
at $\gamma_n$. The input space carries the norm
\begin{equation}
  \|\gred'-\gred\|
  =|b'-b|+\|\ell'-\ell\|_2+\|\Om'-\Om\|_{\mathrm{op}}
   +\|\xired{0}'-\xired{0}\|_2+\|\xired{1}'-\xired{1}\|_2.
  \label{eq:opt-input-norm}
\end{equation}
We need feasible perturbations and convergence of nearly optimal solutions
to differentiate this program. Both follow from relative Slater and the
bounded population lower level set, as the following lemma establishes.

\begin{lemma}[Feasible perturbations and optimizer stability]
\label{lem:optimizer-stability}
Under \cref{ass:regular-endpoint}, the feasible mapping of \eqref{eq:setup-branch-program}
is directionally regular at the population optimizer $\beta^\star$ in
every reduced-input direction. For every sequence
$\gamma_n=\gamma^\ast+t_n\dot\gamma_n$ with $t_n\downarrow0$ and
$\dot\gamma_n\to\dot\gamma$, there are feasible points
$\tilde\beta_n=\beta^\star+O(t_n)$ for the perturbed programs.
Every sequence of $\varepsilon_n$-optimal solutions with
$\varepsilon_n=o(t_n)$ is bounded and converges to $\beta^\star$.
\textnormal{\hyperref[subsec:optimizer-stability-proof]{Proof in Appendix~\ref*{subsec:optimizer-stability-proof}.}}
\end{lemma}

The objective and constraints are smooth in $(\beta,\gamma)$, the
population program is convex, and \cref{ass:regular-endpoint} gives an attained optimum.
\Cref{lem:optimizer-stability} supplies directional regularity and a
convergent sequence of exact optimizers for each input perturbation.
These are the hypotheses of \citet[Theorem~4.24]{bonnans2000perturbation},
which gives the Hadamard directional derivative as the infimum over
population optimizers of the supremum over optimal duals of the
Lagrangian derivative. Under \cref{ass:regular-endpoint}, both sets are singletons.
Consequently, we differentiate \eqref{eq:setup-reduced-lagrangian} at the
unique population KKT triple to obtain
\begin{equation}
  D\branchval(\gred^\ast)[\dot\gred]
  =\dot b+\dot\ell^\top(\beta^\star_1-\beta^\star_0)
   +\sum_{a\in\{0,1\}}\lamred_a^{\star\top}
       (\dxi{a}-\dot\Om\Bspan^\top\beta_a^\star).
  \label{eq:opt-envelope}
\end{equation}
The inequality multiplier contributes no term because $(G,h)$ is fixed.
The expression is continuous and linear in $\dot\gamma$. Since
$\gamma^\ast+t_n\dot\gamma_n=\gamma^\ast+t_n\dot\gamma+o(t_n)$, the
Hadamard directional conclusion gives
\begin{equation}
  \frac{\branchval(\gred^\ast+t_n\dot\gred_n)
             -\branchval(\gred^\ast)}{t_n}
  \longrightarrow D\branchval(\gred^\ast)[\dot\gred]
  \label{eq:opt-hadamard-definition}
\end{equation}
for every convergent sequence of directions in the open reduced domain.
This proves ordinary Hadamard differentiability in those coordinates.

\paragraph{Derivative in the original moments.}
To express the derivative in the original moments, substitute
$\dot\Om=B_C^\top\dot\Sigma B_C$ and
$\dot\varsigma_a=B_C^\top\dot d_a$ into \eqref{eq:opt-envelope}. Because the
chart and its inverse are fixed linear maps, the same sequential limit
holds tangentially on $\mathcal G_C$, with
\begin{equation}
  D\branchval(\Gamma)[\dot\Gamma]
  =\dot b+\dot\ell^\top(\beta_1^\star-\beta_0^\star)
   +\sum_a(B_C\bar\lambda_a^\star)^\top
          (\dot d_a-\dot\Sigma\beta_a^\star).
  \label{eq:opt-site-domain-derivative}
\end{equation}
It remains to check that this expression is independent of the full dual
representative. The reduced KKT conditions are
\begin{align}
  c(\ell)-\Hred(\Om)^\top\lamred+G^\top\nuineq&=0,
  \label{eq:opt-kkt-stationarity}\\
  \Hred(\Om)\beta-\ered&=0,
  \label{eq:opt-kkt-equality}\\
  G\beta\le h,\qquad\nuineq\ge0,\qquad
  \nuineq^\top(G\beta-h)&=0.
  \label{eq:opt-kkt-complementarity}
\end{align}
These conditions are necessary and sufficient by convexity and relative
Slater \citep[Theorems~28.2--28.3]{rockafellar1970convex}. Projecting any
full equality dual onto $\Vspan$ gives the unique reduced dual, so every
full representative is $B_C\bar\lambda_a^\star+z_a$ with
$z_a\in\kernel(\CovC)=\Vspan^\perp$. Orthogonality gives
\[
  \|B_C\bar\lambda_a^\star+z_a\|^2
  =\|B_C\bar\lambda_a^\star\|^2+\|z_a\|^2,
\]
which identifies $B_C\bar\lambda_a^\star$ as the minimum-norm
representative. Also,
\[
  \dot d_a-\dot\Sigma\beta_a^\star
  =B_C(\dot\varsigma_a-\dot\Om B_C^\top\beta_a^\star)\in\Vspan.
\]
Thus $z_a$ contributes zero to \eqref{eq:opt-site-domain-derivative},
proving the claimed invariance.

\paragraph{Upper endpoint.}
For the upper endpoint, let $R$ negate the $\ell$ coordinate and leave
all other reduced coordinates fixed. The feasible set is unchanged, and
\begin{equation}
  v_+(\gamma)=2b-\branchval(R\gamma).
  \label{eq:opt-endpoint-reflection}
\end{equation}
Applying the derivative already proved to the reflected minimization gives
\[
\begin{aligned}
  Dv_+(\gamma^*)[\dot\gamma]
  &=2\dot b-D\branchval(R\gamma^*)[R\dot\gamma]\\
  &=\dot b+\dot\ell^\top(\beta_{+,1}^*-\beta_{+,0}^*)
    -\sum_a\bar\lambda_{+,a}^{*,\mathrm{or}\top}
      (\dot\varsigma_a-\dot\Omega B_C^\top\beta_{+,a}^*).
\end{aligned}
\]
Hence the upper endpoint has the same derivative formula with its own
optimizer and $\bar\lambda_{+,a}^*=-\bar\lambda_{+,a}^{*,\mathrm{or}}$.
The full-dual invariance and minimum-norm claims follow by the same
projection argument.

\paragraph{Finite unions.}
For a finite union we impose the following branch-separation assumption
used in the theorem.
\begin{assumption}[Branch separation for finite unions]
\label{ass:branch-separation}
Each endpoint has a unique winning branch and encoding, with a positive
population value gap to every other feasible branch. Every competing
branch is either locally infeasible or has a locally finite continuous
value.
\end{assumption}
The winning value is continuous by the differentiability just proved.
Discard locally infeasible competitors and let $\Delta>0$ be the smallest
population gap among those remaining. By continuity, on a sufficiently
small common neighborhood every branch value differs from its population
value by less than $\Delta/3$. The winner therefore retains a gap of at
least $\Delta/3$. If no competitors remain, it is the sole feasible branch
throughout a neighborhood. In either case the union value agrees locally
with the winning branch and inherits its derivative. A consistent input
estimate lies in this neighborhood with probability tending to one, so the
union and winning-branch estimators are identical on that event and differ
by $o_p(n^{-1/2})$. This completes the proof of
\cref{prop:value-derivative}.

\begin{remark}[Irregular laws]
\label{rem:irregular}
Uniqueness is needed for the directional derivative to be linear. For
example, $\min_{|x|\le1}tx=-|t|$ has directional derivative $-|q|$ at
$t=0$. At a tie between optimizers or branches, Gaussian inference can
therefore fail unless the tied derivatives agree
\citep{fang2019inference}. \Cref{ass:branch-separation} also imposes local finiteness
and continuity of the competitors separately. Fixed polyhedral
inequalities and full-row-rank equalities alone do not ensure these
properties. For example, with $p=2$, $r=1$, $B_C=(1,0)^\top$, $\Omega=1$,
$\varsigma_0=\varsigma_1=0$, and restrictions $\beta_{0,2},\beta_{1,2}\ge0$, the branch
value at $b=1$ and $\ell=(0,\epsilon)^\top$ is
$\inf_{\beta_{0,2},\beta_{1,2}\ge0}\{1+\epsilon(\beta_{1,2}-\beta_{0,2})\}$.
It equals $1$ at $\epsilon=0$ and $-\infty$ elsewhere.
\end{remark}

\subsubsection{Proof of \texorpdfstring{\hyperref[lem:optimizer-stability]{Lemma~\ref*{lem:optimizer-stability}}}{Optimizer stability}}
\label{subsec:optimizer-stability-proof}
Fix a direction $\dot\gamma$ and write
$\dot H=\Hred(\dot\Omega)$ and $\dot e=(\dot\varsigma_0^\top,\dot\varsigma_1^\top)^\top$.
Full row rank gives a solution to the linearized equality,
\[
  \dot\beta_0=H_0^\top(H_0H_0^\top)^{-1}(\dot e-\dot H\beta^*).
\]
Let $\beta^{\mathrm{sl}}$ be the population Slater point. Adding a
sufficiently large multiple of $\beta^{\mathrm{sl}}-\beta^*$ to
$\dot\beta_0$ leaves the equality unchanged and strictly decreases every
active inequality, because
$H_0(\beta^{\mathrm{sl}}-\beta^*)=0$ and
$G_i(\beta^{\mathrm{sl}}-\beta^*)<0$ for each active $i$. Thus we obtain
$\dot\beta$ with
\begin{equation}
  H_0\dot\beta+\dot H\beta^*-\dot e=0,\qquad
  G_i\dot\beta<0\quad\text{whenever }G_i\beta^*=h_i.
  \label{eq:opt-gollan-condition}
\end{equation}
This is the Gollan condition, which gives directional regularity by
\citet[Proposition~3.4(ii)]{bonnans1998guided} and
\citet[Theorem~4.9]{bonnans2000perturbation}.

For a path $\gamma_n=\gamma^*+t_n\dot\gamma_n$, with
$\dot\gamma_n\to\dot\gamma$, the equality residual
$r_n=H_n(\beta^*+t_n\dot\beta)-e_n$ is $o(t_n)$.
Correct it exactly by setting
\[
  q_n=-H_n^\top(H_nH_n^\top)^{-1}r_n,
  \qquad \tilde\beta_n=\beta^*+t_n\dot\beta+q_n.
\]
The singular-value bound \eqref{eq:opt-sigma-min} gives $q_n=o(t_n)$.
Every active inequality has slack
$t_nG_i\dot\beta+o(t_n)<0$, and inactive inequalities remain strict by
continuity. Thus $\tilde\beta_n$ is feasible and
$\tilde\beta_n=\beta^*+O(t_n)$.

We next show that nearly optimal solutions cannot escape to infinity.
The bounded population lower level set in \cref{ass:regular-endpoint} rules out any
nonzero recession direction, so
\begin{equation}
  \{z:H_0z=0,\ Gz\le0,\ g^\top z\le0\}=\{0\}.
  \label{eq:opt-no-recession-lower}
\end{equation}
Indeed, any such nonzero $z$ would make
$\beta^{\mathrm{sl}}+tz$, $t\ge0$, an unbounded ray in that level set.
Conversely, normalizing an unbounded sequence in the level set and taking
a subsequential limit produces such a nonzero $z$.

Let $\beta_n$ be $\varepsilon_n$-optimal, with
$\varepsilon_n=o(t_n)$. If it were unbounded along a subsequence, set
$\rho_n=\|\beta_n-\beta^{\mathrm{sl}}\|\to\infty$ and pass to a further
subsequence with
$z_n=(\beta_n-\beta^{\mathrm{sl}})/\rho_n\to z$, $\|z\|=1$.
Feasibility and continuity give $H_0z=0$ and $Gz\le0$.
Local boundedness of the value and near-optimality give
$b_n+g_n^\top\beta_n\le M+o(1)$ for some finite $M$.
Dividing by $\rho_n$ and taking limits yields $g^\top z\le0$,
contradicting \eqref{eq:opt-no-recession-lower}. Thus $\{\beta_n\}$ is
bounded.

Every limit point $\beta_\infty$ is population-feasible. The feasible
sequence constructed above satisfies $\tilde\beta_n\to\beta^*$, and
near-optimality gives
\[
  b_n+g_n^\top\beta_n
  \le\branchval(\gamma_n)+\varepsilon_n
  \le b_n+g_n^\top\tilde\beta_n+\varepsilon_n.
\]
Passing to a convergent subsequence shows
$b^*+g^\top\beta_\infty\le b^*+g^\top\beta^*$.
Feasibility gives the reverse inequality, so $\beta_\infty$ is a
population optimizer. Uniqueness makes every limit point equal to
$\beta^*$, proving $\beta_n\to\beta^*$.

\subsection{Proof of \texorpdfstring{\hyperref[lem:reported-variance]{Lemma~\ref*{lem:reported-variance}}}{Variance consistency}}
\label{subsec:reported-variance-proof}
The input expansion gives $\what\gamma\cp\gamma^*$ within the open
reduced domain. To prove consistency of the reported variance, we first
need convergence of the optimizer and dual variables used in the fitted
influence functions.

\begin{proposition}[Consistency of reduced KKT selections]
\label{prop:kkt-consistency}
Assume that \cref{ass:regular-endpoint} holds. Let $\gred_n\to\gred^\ast$ in $\redomain$ and let $\kkt_n=(\beta_n,\lamred_n,\nuineq_n)$ be
\emph{any} measurable selection of KKT triples for $\gred_n$. Then $\kkt_n\to\kkt^\star(\gred^\ast)$.
\textnormal{\hyperref[proof:kkt-consistency]{Proof in Appendix~\ref*{proof:kkt-consistency}.}}
\end{proposition}

For a finite union, \cref{prop:value-derivative} places the reported
solution on the population-winning branch with probability tending to one.
\Cref{prop:kkt-consistency} then gives convergence of the optimizer and
reduced dual. The lift $\lambda_a=B_C\bar\lambda_a$ is fixed and linear,
so the minimum-norm full dual converges as well.

We next separate nuisance replacement from replacement of these global
optimization estimates. Let $\tilde\phi_{j,k}$ denote the fold-$k$ source
influence function using fitted nuisances but population coefficients and
dual variables. Define $\tilde\phi_{T,j,k}$ analogously. In
\eqref{eq:source-if}, the nuisance errors enter through
$\what w\what{\widetilde U}_a-w\widetilde U_a$,
$\what w\what\Ctil-w\Ctil$, and
\[
  \what\Ctil(\what{\widetilde U}_a-\what\Ctil^\top\beta^*_{j,a})
  -\Ctil(\widetilde U_a-\Ctil^\top\beta^*_{j,a}).
\]
Expanding these products gives terms with a nuisance error as one factor
and an $L_4(P)$-bounded quantity as the other. The interpolation argument
in the proof of \cref{prop:primitive-gradients} therefore implies
\[
  \|\tilde\phi_{j,k}-\phi_j\|_{P,2}=o_p(1),\qquad
  \|\tilde\phi_{T,j,k}-\phi_{T,j}\|_{Q_T,2}=o_p(1).
\]
For the target function, this follows directly from target $L_2$
consistency of the marginal regressions. Conditional Markov's inequality
on each holdout now makes the empirical averages of the squared
differences $o_p(1)$. Pooling finitely many folds preserves this conclusion.

The optimizer and dual estimates use the whole sample, so we handle their
replacement by a deterministic bound. At fixed fitted nuisances, the
source influence function is a finite linear combination of
\[
  \what w\what{\widetilde U}_a,\qquad \what w\what\Ctil,\qquad
  \what\Ctil\what{\widetilde U}_a,\qquad
  \operatorname{vec}(\what\Ctil\what\Ctil^\top).
\]
The coefficients are continuous functions of the optimizer and dual
variables. Their changes are $o_p(1)$ by the convergence already proved.
Each displayed fitted quantity has an $O_p(1)$ empirical second moment:
its conditional population second moment is $O_p(1)$ by the $L_q$
bounds, and conditional Markov's inequality transfers this bound to the
holdout. Thus, by Cauchy--Schwarz in this finite-dimensional combination,
replacing the optimization estimates changes the pooled empirical
squared difference by $o_p(1)$. The target argument uses the fitted
marginal regressions in the same way. Estimated centering constants
converge by the input expansion; alternatively, the empirical variance
removes these constants exactly.

Combining the two replacements gives
\[
  \frac1n\sum_i\{\what\phi_j(O_i)-\phi_j(O_i)\}^2=o_p(1),\qquad
  \frac1{n_T}\sum_i
    \{\what\phi_{T,j}(X_{T,i})-\phi_{T,j}(X_{T,i})\}^2=o_p(1).
\]
Here each fitted function uses the nuisance estimates for that observation's
fold. Cauchy--Schwarz and the ordinary law of large numbers for the
population influence functions imply convergence of the empirical first
and second moments. Their centered empirical variances consequently
converge to the two population variances. Multiplication of the target
variance by $n/n_T\to\zeta^{-1}$ proves
$\what V_j\to_p V_j$. For the known-target design the same proof applies
to the source variance alone.

\subsubsection{Proof of \texorpdfstring{\hyperref[prop:kkt-consistency]{Proposition~\ref*{prop:kkt-consistency}}}{Consistency of reduced KKT selections}}
\label{subsec:kkt-consistency}
\label{proof:kkt-consistency}
We first prove $\beta_n\to\beta^\star$. If this failed, there would be a subsequence with
$\|\beta_n-\beta^\star\|\ge\varepsilon$ for some $\varepsilon>0$. Terms with $\gred_n=\gred^\ast$ cannot
occur on this subsequence because the population optimizer is unique. Put
\begin{equation}
  t_n=\|\gred_n-\gred^\ast\|,
  \qquad
  \dot\gred_n=\frac{\gred_n-\gred^\ast}{t_n}.
  \label{eq:opt-normalized-arbitrary-sequence}
\end{equation}
The coordinate space is finite dimensional. After passing to a further subsequence,
$\dot\gred_n\to\dot\gred$ and $t_n\downarrow0$. \Cref{lem:optimizer-stability}, applied with
$\varepsilon_n=0$, then gives $\beta_n\to\beta^\star$, a contradiction. Thus the optimizer convergence
holds for the original sequence.

We next construct a uniformly strict feasible point for the perturbed equality system. We state
the construction for a general input rather than along the sequence. Let $\beta^{\mathrm{sl}}$ be the population Slater point and, for $\gred\in\redomain$,
define
\begin{equation}
  q(\gred)=\Hred(\Om)^\top\bigl(\Hred(\Om)\Hred(\Om)^\top\bigr)^{-1}
  \bigl(\ered-\Hred(\Om)\beta^{\mathrm{sl}}\bigr),\qquad
  \beta^{\mathrm{sl}}(\gred)=\beta^{\mathrm{sl}}+q(\gred).
  \label{eq:opt-perturbed-slater-point}
\end{equation}
Full row rank gives $\Hred(\Om)\beta^{\mathrm{sl}}(\gred)=\ered$. The map $\gred\mapsto q(\gred)$ is
continuous on $\redomain$ and $q(\gred^\ast)=0$, because $\Hred_0\beta^{\mathrm{sl}}=\ered_0$. Since
$G\beta^{\mathrm{sl}}<h$, there are $\sigma>0$ and a neighbourhood $\redomain_{\mathrm{sl}}$ of
$\gred^\ast$ such that
\begin{equation}
  G\beta^{\mathrm{sl}}(\gred)\le h-\sigma\one
  \qquad\text{and}\qquad
  \|\beta^{\mathrm{sl}}(\gred)\|\le M_{\mathrm{sl}}:=\|\beta^{\mathrm{sl}}\|+1
  \qquad\text{for every }\gred\in\redomain_{\mathrm{sl}} .
  \label{eq:opt-uniform-slater-slack}
\end{equation}
Writing $\beta^{\mathrm{sl}}_n=\beta^{\mathrm{sl}}(\gred_n)$, this holds for every $n$ beyond some
$n_0$.

Take the dot product of the stationarity equation $g_n-\Hred_n^\top\lamred_n+G^\top\nuineq_n=0$ with
$\beta^{\mathrm{sl}}_n-\beta_n$. Both points satisfy the equalities at $\gred_n$, and complementary
slackness gives $\nuineq_n^\top G\beta_n=\nuineq_n^\top h$. Hence
\begin{equation}
\begin{split}
  0
  &=g_n^\top(\beta^{\mathrm{sl}}_n-\beta_n)
    +\nuineq_n^\top(G\beta^{\mathrm{sl}}_n-h),\\
  \sigma\|\nuineq_n\|_1
  &\le g_n^\top(\beta^{\mathrm{sl}}_n-\beta_n).
  \label{eq:opt-inequality-multiplier-bound}
\end{split}
\end{equation}
The right side is bounded because $g_n\to g$, $\beta^{\mathrm{sl}}_n\to\beta^{\mathrm{sl}}$, and
$\beta_n\to\beta^\star$. Thus $\{\nuineq_n\}$ is bounded.

Stationarity also gives
\begin{equation}
  \Hred_n^\top\lamred_n=g_n+G^\top\nuineq_n .
  \label{eq:opt-equality-multiplier-bound}
\end{equation}
The smallest singular value of $\Hred_n$ is bounded away from zero by \eqref{eq:opt-sigma-min} and
$\Om_n\to\Om^\ast$, so $\|\lamred_n\|\le\|g_n+G^\top\nuineq_n\|/\sigma_{\min}(\Hred_n)$ and the equality
multipliers are bounded as well.

Take an arbitrary convergent subsequence of the bounded dual sequence. Passing to the limit in
stationarity, primal feasibility, dual feasibility, and complementary slackness shows that its limit is
a population KKT dual pair associated with $\beta^\star$. \Cref{ass:regular-endpoint} makes that pair unique, so the
limit is $(\lamred^\star,\nuineq^\star)$. Every convergent subsequence has the same limit, which proves
$\kkt_n\to\kkt^\star(\gred^\ast)$.

\section{Effect hierarchy as a finite union of linear optimization problems}
\label{app:effect-ordering}
\label{app:mip}

We show that effect hierarchy can be imposed exactly by solving finitely
many linear programs. We then identify how zero direct coefficients can
create tied branches, which prevents the branch-separation condition used
for Gaussian inference.

Partition the site-level variables as $C=(C_D^\top,C_H^\top)^\top$, with
$p_D$ direct terms and $p_H$ higher-order terms, and partition each arm's
coefficients as $\beta_a=(\beta_{a,D}^\top,\beta_{a,H}^\top)^\top$. Effect
hierarchy, restriction \eqref{eq:effect-ordering} of \cref{sec:small-sites},
bounds every higher-order coefficient by a multiple of every direct
coefficient in the same arm. Given a sensitivity level $\kappa\geq0$, it
requires
\begin{equation}
  |\beta_{a,H,j}|\leq\kappa|\beta_{a,D,k}|
  \quad\text{for every }a\in\{0,1\},\ j\leq p_H,\ k\leq p_D.
  \label{eq:eo-restriction}
\end{equation}
The comparison is per arm and over all direct--higher-order pairs. Reporting
guidance appears in \cref{subsec:eo-diagnostics}.

\subsection{Exact representation and local branch selection}
\label{subsec:eo-enumeration}

For each arm $a$ and direct term $k$, choose a sign $\omega_{a,k}\in\{-1,1\}$. On that branch impose
\begin{align}
  \omega_{a,k}\beta_{a,D,k}&\geq0,
  \label{eq:branch-sign}\\
  \beta_{a,H,j}&\leq
  \kappa \omega_{a,k}\beta_{a,D,k},
  \label{eq:branch-eo-upper}\\
  -\beta_{a,H,j}&\leq
  \kappa \omega_{a,k}\beta_{a,D,k}
  \label{eq:branch-eo-lower}
\end{align}
for every higher-order term $j$ and every direct term $k$. Conditional on \eqref{eq:branch-sign}, the
right-hand side equals $\kappa|\beta_{a,D,k}|$. Thus
\eqref{eq:branch-eo-upper}--\eqref{eq:branch-eo-lower} are exactly
$|\beta_{a,H,j}|\leq\kappa|\beta_{a,D,k}|$. Each sign pattern $\omega=(\omega_{a,k})$ therefore contributes one polyhedron
$F(\omega)=\{\beta:G_\omega\beta\le h_\omega\}$ of the finite union \eqref{eq:setup-restriction-union}, with
$(G_\omega,h_\omega)$ fixed by the analyst and not varying with $\gred$, as \cref{ass:regular-endpoint} requires.

Every coefficient vector satisfying effect hierarchy has a compatible
sign pattern, obtained from its direct coefficients; a zero coefficient
may use either sign. Conversely, on any branch,
$\omega_{a,k}\beta_{a,D,k}=|\beta_{a,D,k}|$, so its inequalities imply \eqref{eq:eo-restriction}. Hence the effect-hierarchy set equals the union over all
$2^{2p_D}$ arm-specific direct-sign patterns. Intersecting each branch with
the source normal equations gives polyhedra whose union is exactly the
original feasible set. Enumeration is practical when $p_D$ is small. For each sign pattern, solve the lower and upper linear
optimization problems, discard infeasible branches, and retain both the outer partial identification bound and the second-best
branch value. Recording the branchwise intervals also reveals whether the exact projection set has
gaps.

\subsection{When enumeration creates a tie}
\label{subsec:eo-duplicates}

At a zero direct coefficient, the same optimizer can have two sign
encodings. To see this, orient the endpoint as a minimization, let $\omega$ be
a winning pattern with optimizer $\beta^\star$, and suppose
$\beta^\star_{a,D,k}=0$. The branch inequalities imply
$|\beta^\star_{a,H,j}|\le\kappa|\beta^\star_{a,D,k}|=0$ for every $j$,
so $\beta^\star_{a,H}=0$. Flipping only $\omega_{a,k}$ gives a pattern $\omega'$
whose changed inequalities still hold with equality at $\beta^\star$.
Thus $\beta^\star\in F(\omega')$ and
$\branchval(\omega')\le\branchval(\omega)$. Since $\omega$ attains the winning value,
the two values coincide and $\beta^\star$ is optimal on both branches.
Conversely, feasibility under both signs forces
$\omega_{a,k}\beta_{a,D,k}\ge0$ and $-\omega_{a,k}\beta_{a,D,k}\ge0$, hence
$\beta_{a,D,k}=0$.

The same argument applies when several zero coordinates are flipped.
Thus
\begin{equation}
  \min_{a\in\{0,1\},\,k\le p_D}|\beta^\star_{a,D,k}|>0
  \label{eq:eo-nonzero-winning-direct}
\end{equation}
makes the optimizer feasible under exactly one sign encoding. Local
selection also requires the positive gap to every other feasible branch
and the competitor continuity required by \cref{ass:branch-separation}.

A shared population optimizer does not imply that tied branch value maps
have the same derivative. An arbitrarily small perturbation can select
different branches. Unless their derivatives are proved to agree, such a
tie does not establish ordinary Hadamard differentiability of the union
value. The Gaussian result in \cref{thm:endpoint-clt} uses
\cref{ass:branch-separation} to exclude ties through a unique winning encoding and a positive
gap.

Nothing in this construction is specific to effect hierarchy. Any restriction written as a finite
union of polyhedra with fixed $(G_\omega,h_\omega)$ enters the theory in the same way, through one branch
program per polyhedron together with the checks of
\cref{ass:regular-endpoint,ass:branch-separation}.

\begin{remark}[The per-arm, all-pairs convention]
\label{rem:eo-agree}
Two design choices in \eqref{eq:eo-restriction} deserve emphasis, because natural variants exist. The
comparison is per arm and over all pairs $(j,k)$, so each higher-order coefficient is bounded by
$\kappa$ times the \emph{smallest} direct magnitude of its arm; a parent-specific variant, indexing
$k$ over the parents of interaction $j$ only, is strictly weaker. And the aggregate family
\eqref{eq:l1-budget} is likewise written per arm; writing it on the contrast block
$\beta_1-\beta_0$ with a single index map would produce a different set. The paper adopts the
per-arm, all-pairs forms throughout and the case study says so when interpreting them.
\end{remark}

\section{Experimental and application details}
\label{app:simulation-application-details}

\subsection{Partial identification with two source sites figure}
\label{subsec:id-two-profiles}

This subsection gives the details behind \cref{fig:two-sites}. Take $X$ to be empty and suppose the
treatment-effect surface is
\begin{equation}
  \tau(c)=\tau_0+c_1\theta_1+c_2\theta_2.
\end{equation}
The two source contexts $(0,0)$ and $(1,1)$ identify $\tau_0$ and $\rho=\theta_1+\theta_2$. Centering the
two profiles gives a covariance proportional to
\begin{equation}
  \begin{bmatrix}1&1\\1&1\end{bmatrix},
\end{equation}
whose range is $\operatorname{span}(1,1)$ and whose null space is $\operatorname{span}(1,-1)$. The
affine fiber is $(\theta_1,\theta_2)=(t,\rho-t)$. A target at $(2,2)$ annihilates the null direction and
has identified effect $\tau_0+2\rho$, despite lying outside the convex hull of the source profiles. At
$(1,0)$, a box $|\theta_1|,|\theta_2|\leq M$ gives
\begin{equation}
  \max\{-M,\rho-M\}\leq\theta_1\leq\min\{M,\rho+M\}.
\end{equation}
Therefore
\begin{equation}
  \tau(1,0)\in
  \left[
    \tau_0+\max\{-M,\rho-M\},
    \tau_0+\min\{M,\rho+M\}
  \right].
\end{equation}
The optimization problem is infeasible if $|\rho|>2M$, unbounded without the box, and point identified
for the on-span target. The span, rather than the convex hull, is the relevant geometry under the
linear response model.

\subsection{The standard one-step correction}
\label{app:one-step-benchmark}

The standard one-step correction (\cref{alg:crossfit-onestep}) is the benchmark used in \cref{sec:simulation}. On each fold it solves
the partial identification bound program at a fitted, uncorrected input and adds the held-out mean of
the influence function. Write $\what\Gamma^{\mathrm{pl},(k)}$ for that fitted
input, $\what\xi^{(k)}$ for its optimizer and dual variables, and
$v(\what\Gamma^{\mathrm{pl},(k)})$ for its lower value. Here $n_k$ is the
source fold size, $\Pnk{k}$ and $\mathbb Q_{T,k}$ average over the held-out
source and target folds, and $\phi,\phi_T$ are the influence functions derived in
\cref{subsec:crossfitting}. The lower endpoint is
\begin{equation}
  \vhatE{\Etwo}{-}=\sum_{k=1}^{K}\frac{n_k}{n}
  \Bigl\{v\bigl(\what\Gamma^{\mathrm{pl},(k)}\bigr)
  +\Pnk{k}\,\phi\bigl(\cdot\,;\what\eta_k,\what\xi^{(k)}\bigr)
  +\mathbb Q_{T,k}\,\phi_T\bigl(\cdot\,;\what\eta_k,\what\xi^{(k)},
  v(\what\Gamma^{\mathrm{pl},(k)})\bigr)\Bigr\},
  \label{eq:onestep-estimator}
\end{equation}
with an analogous expression for the upper endpoint.

\begin{algorithm}[htbp]
\caption{Standard one-step correction with foldwise optimization}
\label{alg:crossfit-onestep}
\begin{algorithmic}[1]
\Require the same inputs as our estimator (\cref{alg:crossfit-plugin})
\State Split both samples into $K$ paired balanced folds.
\For{$k=1,\ldots,K$}
  \State Split the rows outside the $k$th folds once more, in both samples,
  into a fitting part $J_{k,\mathrm{fit}}$ and an input part
  $J_{k,\mathrm{in}}$ of proportional sizes.
  \State Fit $\what\eta_k$ on $J_{k,\mathrm{fit}}$ only, fitting joint site
  probabilities so that $C-\what m_{C,k}(X)\in\mathcal V_C$.
  \State Form the fitted inputs $\gplugk{k}$ as sample moments at
  $\what\eta_k$ over $J_{k,\mathrm{in}}$, and express the equality system in
  the $B_C$ coordinates.
  \State Solve the partial identification bound program at $\gplugk{k}$, recording its value,
  optimizer, reduced equality dual, and inequality dual.
  \State Evaluate the source and target influence functions on the held-out
  rows and average each.
\EndFor
\State Record whether every program was feasible. For a union of polyhedra,
record which polyhedron attains each endpoint in each fold.
\State If every fold is feasible and, for each endpoint, the selected polyhedron
is the same across folds, report
\eqref{eq:onestep-estimator}, its upper partial identification bound analogue, and Wald intervals.
Otherwise, declare a fold failure.
\end{algorithmic}
\end{algorithm}

For our estimator (\cref{alg:crossfit-plugin}), the theory requires small products of
nuisance errors and consistency of the optimizer and dual estimates used
for the standard error. For the standard one-step correction (\cref{alg:crossfit-onestep}), those estimates also
enter the point correction, so theory will require stronger rates on the
fitted inputs. It can then be shown that the standard one-step correction is asymptotically equivalent to our estimator and converges to the same limit.

\subsection{Monte Carlo design}
\label{app:simulation-details}

All four simulation designs use the arm-specific site-level model of
\cref{ass:model} with $Y=g_A(X)+C^\top\beta_A+\varepsilon$,
$\varepsilon\sim N(0,1)$, where $g_1=g_0+\tau$ and
$\beta_1=\beta_0+\theta$ in treatment-shift notation. The source law pools $S$ sites with
fixed context profiles $c_s$; each unit draws its site $s$, then
covariates
\[
  X_1\sim\operatorname{Bernoulli}(r_s),\qquad
  X_2\mid s\sim\operatorname{Binomial}(10,q_s),\qquad
  X_3\sim N(0,1)
\]
and treatment with known site-varying probability
$\propen{1}(X,C)\in[0.35,0.65]$. The site-shifted law of $X_2$ (and, in the boosted
designs, of $X_1$) is what makes $\muC$ nontrivial and gives the joint site
classifier something to learn. An independent target covariate sample of
size $n_T=n$ is drawn with $X_1\sim\operatorname{Bernoulli}(1/2)$,
$X_2\sim\operatorname{Binomial}(10,q_T)$, $X_3\sim N(0,1)$, so every run
exercises the two-sample expansion, the density-ratio nuisance, and the
target influence term. Because $n_T=n$ in every design, the sampling ratio
is $\zetaratio=1$ throughout, and the target term of
\eqref{eq:setup-master-variance} enters with weight one.

In the \emph{linear} designs, $g_0$ and $\tau$ are linear in
$(X_1,X_2,X_3)$ and the site mix depends on $X_2$ alone; the nuisances are
fitted by saturated-in-$X_2$ regressions and cell frequencies, which are
correctly specified. In the \emph{boosted} designs, $g_0$ and $\tau$
contain indicator and interaction terms in all three covariates
(for example $\tau(X)=2+1.5\,\one\{X_2\geq6\}-X_1\one\{X_3>0\}
+0.6(X_3^2-1)$), the site mix depends on $(X_1,X_2)$, and every nuisance is
fitted with gradient-boosted trees (histogram gradient boosting, 150
trees, depth 3, $\ell_2$ regularization 1).

\paragraph{Point-identified designs.}
$S=6$ sites face a basis with $p_D=3$ direct terms and one higher-order
product term, $p=4$. The centered site profiles span $\R^4$, so $r=p$ and
every target context difference satisfies \eqref{eq:point-id-condition}. No coefficient
restriction is imposed. The target profile lies outside the convex hull of
the source profiles; its effect is identified through the span, as in
\cref{fig:two-sites}.

\paragraph{Partially identified designs.}
$S=4$ sites face a basis with $p_D=3$ direct terms and $p_H=3$
higher-order terms ($D_1^2$, $D_1D_2$, $D_2D_3$), $p=6$. The centered span
has rank $r=3$, so three directions are unidentified: the few-sites
mechanism of the application, not an artificial duplication of columns.
The restriction set is effect hierarchy at $\kappa=1$, imposed through its
$2^{2p_D}=64$ exact sign branches, intersected with the coefficient box
$|\beta_{a,j}|\leq2.5$; the true coefficients satisfy every restriction
strictly. The population programs satisfy \cref{ass:regular-endpoint}, with a unique
nondegenerate optimal vertex and unique row-reduced dual at each partial identification bound, and they satisfy
\cref{ass:branch-separation} with
branch gaps $0.58$ (lower) and $1.39$ (upper). The smallest absolute direct coefficient at the winning
optimizer is $1.5692$ for the lower partial identification bound and $1.7006$ for the upper partial identification bound, so
the nonzero-encoding check \eqref{eq:eo-nonzero-winning-direct} also holds.

\paragraph{Exact truths.}
Because $X_2$ has finite support and the site mix is known,
$(\CovC,\CovR{0},\CovR{1},\ellvec)$ are exact finite sums, and $\bT$
combines those sums with an 80-node Gauss--Hermite quadrature over $X_3$.
The population partial identification bounds solve the same optimization problems at the exact inputs. The
point-linear and point-boost truths are single values; the partial-linear truth is
$[-3.837,7.869]$ and the partial-boost truth is $[-3.022,8.684]$.

\paragraph{Estimators and reported statistics.}
Every design uses four-fold paired cross-fitting and the rank-preserving
joint site classifier of \cref{subsec:rank-preservation}, so each fitted residual
$C-\what\muC(X)$ lies in $\Vspan$ exactly. Each replication computes the direct plug-in, our estimator (\cref{alg:crossfit-plugin}), and
the standard one-step correction (\cref{alg:crossfit-onestep}) on the same data and outer folds. The sample-size grids are
\begin{align*}
  \mathcal N_{\mathrm{linear}}
  &=\{1{,}000,2{,}500,5{,}000,10{,}000,20{,}000,50{,}000\},\qquad R=300,\\
  \mathcal N_{\mathrm{boosted}}
  &=\{500,1{,}000,2{,}500,5{,}000,10{,}000\},\qquad R=100.
\end{align*}
Saturated linear regressions are fit by minimum-norm least squares, and all
4,600 matched replications completed. Per-replication records include the
endpoints, standard errors, active branches, branch gaps, span diagnostics,
eigenvalues, and solver status.

\subsection{Conditional coverage and sign mismatch}
\label{app:onestep-simulations}
\label{app:matched-simulations}

\Cref{fig:sim-matched-boosted} in the main text reports sign mismatch probability and
average pointwise coverage for the partially identified boosted design.
Coverage for the standard one-step correction is conditional on no sign mismatch for either partial identification bound.

The standard one-step correction had no numerical or fold-level feasibility failures. Every
declared failure was a sign mismatch across folds. In the partial-linear
design, its lower and upper endpoint failure rates were 18.7 and 5.0 percent
at $n=1{,}000$, falling to 1.0 and 0 percent at $n=2{,}500$ and to zero
thereafter. In the partial-boost design, the rates were 78 and 45 percent at
$n=500$, 28 and 19 percent at $n=1{,}000$, 9 and 2 percent at
$n=2{,}500$, and zero from $n=5{,}000$. Our estimator permitted inference
in every replication.

\subsection{Variance matching and difference between the corrections}
\label{app:variance-matching}

At the larger sample sizes, the mean standard errors are close to the
Monte Carlo standard deviations for both algorithms. The main mismatches occur
for the standard one-step correction in the smallest partial designs, where the reported results
are conditional on no sign mismatch for the corresponding partial identification bound.

\begin{figure}[t]
  \centering
  \includegraphics[width=\textwidth]{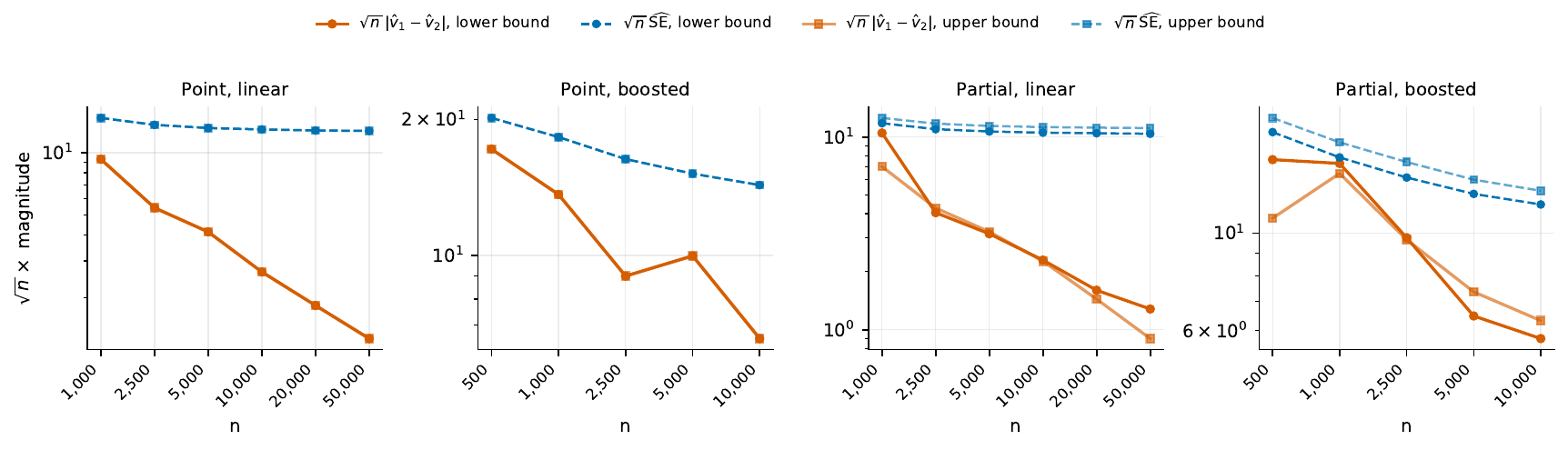}
  \caption{Finite-sample gap between our estimator (\cref{alg:crossfit-plugin})
  and the standard one-step correction (\cref{alg:crossfit-onestep}). Solid curves show
  the mean of
  $\sqrt n\,|\vhatE{\Eone}{j}-\vhatE{\Etwo}{j}|$. Dashed curves show the mean
  value of our estimator's $\sqrt n\,\widehat{\mathrm{SE}}_j$. Means use the
  replications without sign mismatch for either partial identification bound. The gap falls
  overall as sample size grows. At the largest sample size, it is about
  one-tenth of a standard error with linear nuisance fits and one-half of a
  standard error with boosted nuisance fits.}
  \label{fig:sim-matched-gap}
\end{figure}

The standard one-step correction corrects values computed at foldwise fitted inputs, whereas
our estimator evaluates the partial identification bound after it pools corrected inputs.
\Cref{fig:sim-matched-gap} compares their absolute endpoint difference
with our estimator's standard error. The scaled difference falls overall as
sample size grows, especially with linear nuisance fits. At the largest sample
size it is 0.099 of a standard error in the point-linear design and 0.123 and
0.081 for the two partial-linear endpoints. The corresponding ratios are
0.457 in the point-boost design and 0.495 and 0.507 in the partial-boost
design. The linear experiments already have small gaps on this scale, while
the boosted experiments remain visibly different on this sample grid.

\FloatBarrier

\subsection{Practical implementation and sensitivity checks}
\label{subsec:implementation-diagnostics}

In addition to the robustness checks reccomended in \cref{subsec:robustness}, we give additional best practices for implementation here. Restrictions should be chosen before the target outcomes are inspected and
applied as a sequence of sensitivity checks. First fix the context basis and
scaling, which determine what a coefficient restriction means. State the
substantive reason for each restriction and add restrictions one at a time so
readers can see what changes the partial identification bounds. When a restriction depends on a value
such as $M$ or $\kappa$, report results over a reasonable
range. If the restriction sets are nested, the partial identification bounds should also be nested. A
failure of nesting indicates an implementation error. Vary one ingredient at a
time, including the target profile and the set of source sites. An infeasible result
means that the restriction conflicts with the estimated source moments within
the chosen model. An unbounded result means that the restriction leaves a
direction that affects the target uncontrolled.

Some restrictions, including effect hierarchy, are finite unions of branches.
Solve every branch separately and retain its interval. Report the overall
endpoints together with these intervals so that any gaps in the identified set
remain visible. For each endpoint, report the selected branch and its estimated
gap from the closest competing branch. This gap is a useful diagnostic, but it
does not prove that the population branch is separated. A small gap or a tie
means that branch selection may be unstable and the usual normal approximation
may fail. \Cref{prop:value-derivative} gives the formal result when the
population gap is positive. The standard one-step correction also requires every fold program to be
feasible and every fold to select the same branch for a given endpoint. If the
folds disagree, report a failed endpoint rather than combining corrections
from different branches.

Numerical checks are needed because the optimizer and reduced equality dual
enter the influence function. The program should be solved in the reduced site
span so that redundant rows of the full equality system do not create artificial dual
nonuniqueness. For every endpoint, check primal and dual feasibility. Then
check stationarity and complementarity, and confirm that the objective gap is
within tolerance. Report the numerical rank within $\mathcal V_C$, the
smallest reduced eigenvalue, and the dual uniqueness check. The error in the reported value should be
negligible relative to $n^{-1/2}$. For our estimator, the optimizer and dual
affect the standard error but not the point estimate. For the standard one-step correction, they
also enter the point correction, so the accuracy requirements are tighter. Do
not replace an infeasible or unbounded result with a clipped finite value.
Record the status and include it in the sensitivity analysis.

\subsection{Multipliers, branch stability, and reporting}
\label{subsec:eo-diagnostics}

\paragraph{Multiplier structure.}
The inequality multiplier $\nuineq^\star$ does not enter
\eqref{eq:opt-envelope}, because $(G,h)$ is fixed along moment
perturbations. The proof in \cref{subsec:proof-proposition-two} accounts
for the changing equality constraints when differentiating the value.
The inequality multiplier still determines the equality dual through
stationarity \eqref{eq:opt-kkt-stationarity} and hence affects the
influence function through $\lamred^\star$.

\paragraph{Branch stability and transfer.}
\Cref{prop:value-derivative} covers both the local constancy of the selected branch and the
$o_p(n^{-1/2})$ transfer from the mixed-integer program to the selected convex branch, so no separate
selection or transfer lemma is needed. For effect hierarchy,
\cref{ass:branch-separation} requires the unique winning encoding and positive gap described in
\cref{subsec:eo-duplicates}.

\paragraph{Reporting.}
Report the estimated branch gap and the dual-uniqueness check
as solver diagnostics, and state explicitly that conditioning
inference on a reported gap is a pretest rather than a certificate of regularity. Report a path over
$\kappa$ rather than treating effect hierarchy as a maintained scientific assumption.

\label{subsec:eo-checks}
Two further checks catch many remaining errors in the partial identification bound calculation. The
rank report and the handling of infeasible and unbounded results are already
covered in \cref{subsec:implementation-diagnostics}.

\begin{enumerate}
  \item Check $\|\what\CovC\what\beta_a-\what{\dmom{a}}\|$ against the solver tolerance for every
  reported branch.
  \item For nested values of $\kappa$, verify that the reported outer intervals are nested in the
  direction implied by the restriction sets.
\end{enumerate}

\subsection{Target sampling and experimental designs}
\label{subsec:supp-designs}

The expansion in \eqref{eq:asymptotic-linearity} treats the source and target
covariate samples as independent. If $\tP$ is known from a census or another
effectively deterministic population file, its influence term is omitted. A
finite target survey contributes the second term in
\eqref{eq:asymptotic-linearity}. Treating such a survey as fixed without a
separate conditional-inference argument understates uncertainty.

Site-specific individual randomization is allowed through $\propen{a}(X,C)$, and
the simulations exercise site-varying probabilities. If treatment is
randomized at some within-site cluster (e.g., by village), $\pseudo{a}$ must use the cluster design
probability and variance estimation must aggregate the influence-function
values at the independent randomization unit. The individual-i.i.d. theorem
does not justify a cluster-robust application by itself; the application's
standard errors sum influence values within each country's randomization
unit as a design-based
correction, and a formal
clustered version of \cref{thm:setup-master} is left to future work.

\end{document}